\documentclass[12pt]{article}
\evensidemargin=\oddsidemargin
  \let\LARGE=\large
  
  \let\large=\normalsize
  \let\ps=\psi
  \newcommand{\beq}{\begin{equation}}
  \newcommand{\eeq}{\end{equation}}
  \newcommand{\beql}[1]{\begin{equation}\label{eq:#1}}

  \newcommand{\beqa}{\begin{eqnarray}}
  \newcommand{\eeqa}{\end{eqnarray}}
  \newcommand{\beqas}{\begin{eqnarray*}}
  \newcommand{\eeqas}{\end{eqnarray*}}
  \newtheorem{Theorem}{Theorem}[section]
  \newtheorem{Proposition}[Theorem]{Proposition}

  \newenvironment{Proof}{\begin{trivlist}
    \item[\hskip \labelsep {\em \indent Proof.}]}{\qed\end{trivlist}}
 \newcommand{\qed}{{\em QED}}
  \newcommand{\B}{{\bf B}}
  \newcommand{\C}{{\bf C}}

  \newcommand{\Q}{{\bf Q}}
  \newcommand{\R}{{\bf R}}

  \newcommand{\al}{\alpha}
  
  \newcommand{\ch}{\chi}
  \newcommand{\da}{\dagger}
  
  \newcommand{\de}{\delta}
  \newcommand{\ep}{\varepsilon}
  \newcommand{\et}{\eta}

  \newcommand{\la}{\lambda}
  \newcommand{\mb}{\mbox}

  \newcommand{\ph}{\phi}
 \newcommand{\rh}{\rho}
  \newcommand{\si}{\sigma}
  \newcommand{\ta}{\tau}

  \newcommand{\tc}{\tau c}

  \newcommand{\De}{\Delta}
  \newcommand{\Ga}{\Gamma}
  \newcommand{\Om}{\Omega}
  \newcommand{\Ph}{\Phi}
  \newcommand{\Ps}{\Psi}

  \newcommand{\And}{\wedge}

  \newcommand{\Eq}[1]{Eq.~(\ref{eq:#1})}

  \newcommand{\Iff}{\Leftrightarrow}

  \newcommand{\Then}{\Rightarrow}
  
  \newcommand{\Tr}{\mbox{\rm Tr}}

  \newcommand{\beqan}{\begin{eqnarray*}}
  \newcommand{\beqar}[1]{\begin{equation}\label{#1}\begin{array}{l}}
  
  \newcommand{\bx}{{\bf x}}
  \newcommand{\by}{{\bf y}}

  \newcommand{\dom}{\mbox{\rm dom}}
  
  \newcommand{\eeqar}{\end{array}\end{equation}}
  \newcommand{\eq}[1]{(\ref{eq:#1})}

\newcommand{\bra}[1]{\langle#1|}
\newcommand{\ket}[1]{|#1\rangle}
\newcommand{\ketbra}[1]{\ket{#1}\bra{#1}}

\newcommand{\bracket}[1]{\langle#1\rangle}
  
  \newcommand{\ran}{\mbox{\rm ran}}

\newcommand{\bmat}{\left[\begin{array}{rr}}
\newcommand{\emat}{\end{array}\right]}
\newcommand{\bvec}{\left[\begin{array}{r}}
\newcommand{\evec}{\end{array}\right]}

  \newcommand{\bA}{{\bf A}}

  \newcommand{\bD}{{\bf D}}

  \newcommand{\bM}{{\bf M}}

  \newcommand{\bP}{{\bf P}}
  
  \newcommand{\bS}{{\bf S}}

  \newcommand{\bZ}{{\bf Z}}
  
  \newcommand{\cB}{{\cal B}}
  \newcommand{\cC}{{\cal C}}

  \newcommand{\cH}{{\cal H}}
  \newcommand{\cI}{{\cal I}}
  
  \newcommand{\cK}{{\cal K}}
  \newcommand{\cL}{{\cal L}}

  \newcommand{\cP}{{\cal P}}

  \newcommand{\cS}{{\cal S}}

\newcommand{\M}{\bM}
\renewcommand{\Re}{\mb{\rm Re}}

\newcommand{\id}{\mb{\rm id}}
\newcommand{\com}{\mb{\rm com}}

\newcommand{\val}[1]{[\![#1]\!]}

\begin{document}

\title{\LARGE\bf{Quantum perfect correlations}\thanks{
A preliminary account on this subject has appeared
in M. Ozawa, {\it Phys.\ Lett.\ A} {\bf 335}, 11-19 (2005).}}
\author{\sc Masanao Ozawa\\
\small\it Graduate School of Information Sciences,
T\^{o}hoku University\\
\small\it Aoba-ku, Sendai,  980-8579, Japan}
\date{}

\maketitle

%%%%% ABSTRACT %%%%%
\begin{abstract}
The notion of perfect correlations between arbitrary observables,
or more generally arbitrary POVMs, is introduced in the standard
formulation of quantum mechanics, and characterized by several 
well-established statistical conditions.  
The transitivity of perfect correlations is proved to generally hold,
and applied to a simple articulation for the failure of Hardy's
nonlocality proof for maximally entangled states.
The notion of perfect correlations between observables and POVMs
is used for defining the notion of a precise measurement of a given
observable in a given state.   A longstanding misconception on
the correlation made by the measuring interaction is resolved
in the light of the new theory of quantum perfect correlations.
\end{abstract}

% TEXT:
\newpage
\section{Introduction}\label{se:intro}

It is often stressed that quantum mechanics does not speak of 
the value of an observable in a single event, but only speaks of 
the average value over a large number of events.
In fact, quantum states are characterized as what determine the
expectation values of all the observables.
However, quantum correlations definitely describe 
relations of values of observables in a single event as typically
in the EPR correlation \cite{EPR35}.
In the early days of quantum mechanics,
the quantum correlation played a central role in measurement
theory since von Neumann \cite{vN32} generally described a process 
of making a perfect correlation between two systems.
In the recent investigations on quantum information \cite{NC00},
the notion of quantum correlations naturally plays a key role,
as in classical information theory the amount of information is 
defined as a
measure of statistical correlations for pairs of random variables.
Nevertheless, we have not had a general notion of quantum 
correlation; in those investigations the quantum correlation 
has rather replaced by the notion of entanglement, 
which is regarded as quantum correlations restricted 
to those between commuting observables from different subsystems.

The main aim of this paper is to establish the general notion of quantum 
perfect correlations.
It should be stressed that statistical correlation is a state dependent
notion, and it is required to address the problem as to when a pair of
observables are considered to be perfectly correlated in a given state. 
The operational meaning of this condition is that those two observables 
can be jointly measured in that state 
and that each joint measurement gives the
same value,  although the value may distribute randomly. 
In classical probability theory,
it is well accepted that two random variables (observables)
are perfectly correlated 
if and only if the joint probability of any pair of their different values 
vanishes.
Thus, we can immediately generalize this notion to pairs of commuting
observables based on the well-defined joint probability distribution 
of commuting observables.
It is well-known that every entangled (pure) state of a bipartite system has 
the Schmidt decomposition that determines naturally 
a pair of perfectly correlated observables in respective subsystems.
The perfect correlation relevant to the study of entanglement
is as such always those for commuting observables.
Nevertheless, we have several problems that strongly demand the 
generalization of the notion of perfect correlations to noncommuting 
observables. 

One of them is the transitivity problem of quantum perfect correlations.
Suppose that commuting 
observables $X$ and $Y$ are perfectly correlated as well as
commuting observables $Y$ and $Z$.  If $X$ and $Z$
commute, we can easily say that $X$ and $Z$ are perfectly
correlated.  However, there are cases where $X$ and $Z$ do not
commute, and no existing theory determines whether $X$ and
$Z$ are considered to be perfectly correlated.

There has been a longstanding misconception on statistical 
correlation in measurement.
In the conventional model of measurement found by von Neumann
\cite{vN32}, the measuring interaction is required to 
establish two different
kinds of perfect correlations: one is between measured observable
{\em before} the interaction and the meter observable {\em after} the
interaction,  and the other is between the meter observable {\em after} the
interaction  and the measured observable {\em after} the interaction.
The first one ensures that the observation of the meter observable
suffices to know the value of the measured observable before the interaction,
and the second one ensures that the measurement leaves the
measured system in the eigenstate pertaining to the measurement result.
However, we have been able to treat only the second correlation,
since the Heisenberg operator of the measured observable before
the interaction and the Heisenberg operator of the meter observable
after the interaction do not commute in general.
Moreover, there has been a confusion between the meaning of 
those two different correlations.
Even in the modern approach to measurement theory, the lack of
the general theory of quantum perfect correlations has left the fundamental 
question unanswered as to when the given observable is precisely 
measured in a given state.

This paper introduces the notion of perfect correlations 
between arbitrary two observables, and characterizes it
by various statistical notions in quantum mechanics.
As a result, the above problems are shown to be answered
by simple and well-founded conditions
in the standard formalism of quantum mechanics.

In Section \ref{se:2}, we introduce the definition of the perfect correlation
between two observables.
In Section \ref{se:3}, the condition that two observables are perfectly
correlated in a given state is characterized in terms of well-formulated
statistical notions in the standard quantum mechanics. 
It is immediate from the definition that two perfectly correlated 
observables are identically distributed, 
i.e., having the same probability distribution, 
but the converse is not true as seen from the case of two 
independent observables with identical distribution.
This section considers the question 
as to what additional condition ensures that two identically distributed
observables are perfectly correlated.

In Section \ref{se:5}, we prove that the perfect correlation between 
observables
in a given state is transitive and consequently is an equivalence relation 
between
observables.
In Section \ref{se:6}, we consider the joint probability distribution 
of perfectly
correlated observables, and show that two observables are perfectly 
correlated if
and only if they have joint  probability distribution concentrated 
on the diagonal. 
We show that our definition of perfectly correlated observables
in a given state ensures that they are jointly measurable in that state.
We characterize the quasi-joint probability distribution 
of perfectly correlated observables.
In Section \ref{se:7}, we consider the perfect correlation between observables 
in bipartite
systems, and characterize pairs of perfectly correlated observables from two 
subsystems.  We also apply the transitivity of perfect correlations to a simple
explanation for the failure of Hardy's nonlocality proof for the class 
of maximally
entangled states.

In Section \ref{se:4}, we consider the perfect correlations between 
probability operator valued measures (POVMs). 
We show that
any pair of POVMs has a joint dilation to a pair of observables 
in an extended
system in such a way that the given POVMs are perfectly correlated 
if and only if the corresponding observables are perfectly correlated.  
In this way, the problem of perfect correlations between POVMs can be
reduced to the problem of  perfect  correlations between observables, and we
extend the characterization of perfectly correlations between two observables
to those between a POVM and an observable.

In Section \ref{se:8}, we consider perfect correlations in measurements, and
gives the definition for precise measurements of an observable 
in a given state, using the notion of perfect correlations between 
observables
and POVMs.
A longstanding misconception on
the correlation made by the measuring interaction is resolved
in the light of the new theory of quantum perfect correlations.
Section \ref{se:9} concludes the present paper with summary and some remarks.

\section{Basic formulations}\label{se:2}

Let $\cH$ be a separable Hilbert space.
An {\em observable} is a self-adjoint operator 
densely defined in $\cH$ and a {\em state} is a
density operator $\rh$ on $\cH$, or equivalently a positive operator $\rh$ 
on $\cH$ with unit trace \cite{vN32}. 
A unit vector $\ps$ in $\cH$ is called a {\em state vector}
or a {\em vector state} defining the state $\rh=\ket{\ps}\bra{\ps}$
that is an extreme point ({\em pure state}) 
in the convex set $\cS(\cH)$ of states on
$\cH$.  Denote by $\cB(\R^{n})$ the Borel $\si$-field of
the Euclidean space $\R^{n}$  
and by $B(\R^{n})$ the algebra of (complex-valued)
bounded Borel functions on $\R^{n}$.
Denote by $\cL(\cH)$ the algebra of bounded operators on $\cH$
and by $\cL(\cH)_{+}$ the cone of positive operators on $\cH$.
A {\em positive operator valued measure} \cite{Ber66} 
is a mapping $\Pi$ from $\cB(\R)$ to $\cL(\cH)_{+}$ such that  
$\Pi(\bigcup_{j}\De_{j})=
\sum_{j=1}^{\infty}\Pi(\De_{j})$ 
in the weak operator topology 
for any disjoint sequence of Borel sets $\De_{1},\De_{2},\ldots$.
A {\em probability operator valued measure (POVM)} \cite{Hel76,Hol82}
is a positive operator valued measure $\Pi$ such that $\Pi(\R)=I$.

We say that two POVMs $\Pi_{1}$ and $\Pi_{2}$ are {\em perfectly correlated} 
in a state $\rh$, iff 
\beqa\label{eq:definition}
\Tr[\Pi_{1}(\De)\Pi_{2}(\Ga)\rh]=0
\eeqa
for any disjoint Borel sets $\De,\Ga$.
For any vector state $\ps$, \Eq{definition} is
equivalent to 
\beqa\label{eq:definition_vector}
\bracket{\Pi_{1}(\De)\ps,\Pi_{2}(\Ga)\ps}=0.
\eeqa
The following proposition generalizes \Eq{definition} to arbitrary pairs of
Borel sets $\De,\Ga$.  

\begin{Proposition}\label{th:equivalence-POVM}
For any POVMs $\Pi_{1},\Pi_{2}$
and any state $\rh$, the following conditions are equivalent.

(i) $\Pi_{1}$ and $\Pi_{2}$ are perfectly correlated in $\rh$.

(ii) $\Tr[\Pi_{1}(\De)\Pi_{2}(\Ga)\rh]=\Tr[\Pi_{1}(\De\cap\Ga)\rh]$ 
for any $\De,\Ga\in\cB(\R)$.

(iii) $\Tr[\Pi_{1}(\De)\Pi_{2}(\Ga)\rh]=\Tr[\Pi_{2}(\De\cap\Ga)\rh]$
for any $\De,\Ga\in\cB(\R)$.

\end{Proposition}
\begin{Proof}
If $\Pi_{1},\Pi_{2}$ are perfectly correlated, we have
\beqas\label{eq:PC-2} 
\Tr[\Pi_{1}(\De)\Pi_{2}(\Ga)\rh]
&=&
\Tr[\Pi_{1}(\De\cap\Ga)\Pi_{2}(\Ga)\rh]
+\Tr[\Pi_{1}(\De\setminus\Ga)\Pi_{2}(\Ga)\rh]\\
&=&\Tr[\Pi_{1}(\De\cap\Ga)\Pi_{2}(\Ga)\rh]\\
&=&\Tr[\Pi_{1}(\De\cap\Ga)\Pi_{2}(\R\setminus\Ga)\rh]
+\Tr[\Pi_{1}(\De\cap\Ga)\Pi_{2}(\Ga)\rh]\\
&=&
\Tr[\Pi_{1}(\De\cap\Ga)\rh]
\eeqas
for any $\De,\Ga\in\cB(\R)$.
This proves (i)$\Then$(ii).
The converse part (ii)$\Then$(i) is obvious, and 
the equivalence (i)$\Iff$(iii) can be proved analogously.
\end{Proof}

Let $\Pi$ be a positive operator valued measure.
For any Borel function $f$ on $\R$
the operator $\Pi(f)$ is defined by 
\beqas
\dom(\Pi(f))&=&\left\{\ps\in\cH\left|
\int_{\R}|f(x)|^{2}\bracket{\ps,d\Pi(x)\ps}<\infty\right\}\right.,\\
\bracket{\ps',\Pi(f)\ps}&=&\int_{\R}f(x)\bracket{\ps',\,d\Pi(x)\ps}
\eeqas
for all $\ps\in\dom(\Pi(f))$ and $\ps'\in\cH$; see \cite{LPY99}
for comparison with other approaches.
For the identity function $\id$ on $\R$, i.e., $\id(x)=x$ for all $x\in\R$,
the operator $\Pi(\id^{n})$ is called the $n$-th {\em moment operator}
of $\Pi$.  For any real-valued Borel function $f$ on $\R$, the relation
\beqa
\Pi^{f}(\De)=\Pi(f^{-1}(\De)),
\eeqa
where $\De\in\cB(\R)$, defines a unique positive operator valued measure $\Pi^{f}$.
For any real-valued Borel functions $f,g$, 
it is easy to see that $\Pi(f\circ g)=\Pi^{g}(f)=\Pi^{f\circ g}(\id)$,
where $f\circ g$ is the composition of $f$ and $g$, i.e., 
$f\circ g(x)=f(g(x))$ for all $x\in\R$.
For any bounded operator $A$ on $\cH$, the relation
\beqa
\Pi^{A}(\De)=A^{\da}\Pi(\De)A,
\eeqa
where $\De\in\cB(\R)$, defines a unique positive operator
valued measure $\Pi^{A}$.
For any bounded operator $A,B$, we have $\Pi^{AB}=(\Pi^{A})^{B}$.
If $\Pi$ is a POVM, so are $\Pi^{f}$ and $\Pi^{U}$
whenever $U$ is isometry.

Now we have the following.
\begin{Theorem}\label{th:equivalence-2}
For any POVMs $\Pi_{1},\Pi_{2}$, state $\rh$,
and unitary operator $U$ on $\cH$,
the following conditions are equivalent.

(i) $\Pi_{1}$ and $\Pi_{2}$ are perfectly correlated in $\rh$.

(ii) $\Pi_{1}^{f}$ and $\Pi_{2}^{f}$ are perfectly correlated in $\rh$ 
for any real-valued Borel function $f$.

(iii) $\Pi_{1}^{f}$ and $\Pi_{2}^{f}$ are perfectly correlated in $\rh$ 
for any bounded real-valued Borel function $f$.

(iv) $\Pi_{1}^{f}$ and $\Pi_{2}^{f}$ are perfectly correlated in $\rh$ 
for a bijective Borel function $f$ from $\R$ to a Borel set $\Om\in\cB(\R)$.

(v) $\Pi^{U}_{1}$ and $\Pi^{U}_{2}$ are perfectly correlated in $U^{\da}\rh U$.
\end{Theorem}
\begin{Proof}
Suppose that $\Pi_{1}$ and $\Pi_{2}$ are perfectly correlated in $\rh$.
Let $f$ be a real-valued Borel function.  We have
\beqas
\Tr[\Pi_{1}^{f}(\De)\Pi_{2}^{f}(\Ga)\rh]
&=&
\Tr[\Pi_{1}(f^{-1}(\De))\Pi_{2}(f^{-1}(\Ga))\rh]
=
\Tr[\Pi_{1}(f^{-1}(\De)\cap f^{-1}(\Ga))\rh]\\
&=&
\Tr[\Pi_{1}(f^{-1}(\De\cap\Ga))\rh]
=
\Tr[\Pi_{1}^{f}(\De\cap\Ga)\rh].
\eeqas
Thus, $\Pi_{1}^{f}$ and $\Pi_{2}^{f}$ are perfectly correlated in $\rh$.
This proves (i)$\Then$(ii).
The implications (ii)$\Then$(iii)$\Then$(iv) are obvious.
Suppose (iv).
Then, there is a Borel function $g$ such that $g[f(x)]=x$ for all $x\in\R$.
By the implication (i)$\Then$(ii), two 
POVMs $\Pi_{1}=\Pi_{1}^{g\circ f}$ and $\Pi_{2}=\Pi_{2}^{g\circ f}$
are perfectly correlated in $\rh$.
This proves (iv)$\Then$(i).  
The equivalence (i)$\Iff$(v) is
straightforward from the property of trace, and the proof is
completed.
\end{Proof}

Let $X$ be an observable on $\cH$.
The {\em spectral measure} of $X$ is the projection-valued POVM
$E^{X}$ such that $E^{X}(p)=p(X)$ for any polynomial $p$.
For any Borel function $f$ on $\R$, the operator $f(X)$ is defined 
by $f(X)=E^{X}(f)$.

We say that two observables $X$ and $Y$ are {\em perfectly correlated} 
in a state $\rh$, iff $E^{X}$ and $E^{Y}$ are perfectly correlated in $\rh$.
From Proposition \ref{th:equivalence-POVM},  
$X$ and $Y$ are perfectly correlated 
in $\rh$ if and only if one of the following equivalent conditions holds:

(i)  $\Tr[E^{X}(\De)E^{Y}(\Ga)\rh]=0$ for any disjoint Borel sets 
$\De,\Ga\in\cB(\R)$.

(ii) $\Tr[E^{X}(\De)E^{Y}(\Ga)\rh]=\Tr[E^{X}(\De\cap\Ga)\rh]$
for any $\De,\Ga\in\cB(\R)$.

(iii) $\Tr[E^{X}(\De)E^{Y}(\Ga)\rh]=
\Tr[E^{Y}(\De\cap\Ga)\rh]$
for any $\De,\Ga\in\cB(\R)$.

The following theorem restates Theorem \ref{th:equivalence-2}
for observables.
\begin{Theorem}\label{th:equivalence-3}
For any observables $X,Y$, state $\rh$,
and unitary operator $U$ on $\cH$,
the following conditions are all equivalent.

(i) $X$ and $Y$ are perfectly correlated in $\rh$.

(ii) $f(X)$ and $f(Y)$ are perfectly correlated in $\rh$ 
for any real-valued Borel function $f$.

(iii) $f(X)$ and $f(Y)$ are perfectly correlated in $\rh$ 
for any bounded real-valued Borel function $f$.

(iv) $f(X)$ and $f(Y)$ are perfectly correlated in $\rh$ 
for a bijective Borel function $f$ from $\R$ to a Borel set $\Om\in\cB(\R)$.

(v) $U^{\da}XU$ and $U^{\da}YU$ are perfectly correlated in
$U^{\da}\rh U$.
\end{Theorem}

From the above theorem, the perfect correlation between two not necessarily
bounded observables $X$ and $Y$ can be reduced to the perfect correlation
of a pair of bounded observables, say, $\tan^{-1}X$ and $\tan^{-1}Y$.

\section{Characterizations of perfectly correlated observables}
\label{se:3}

The {\em cyclic subspace}  of $\cH$ spanned 
by an observable $X$ and a state vector $\ps\in\cH$ 
is the closed subspace $\cC(X,\ps)$ defined by
$$
\cC(X,\ps)=\mb{the closure of }\{f(X)\ps\in\cH\mid f\in B(\R)\}.
$$
Denote by $\cC_{1}(X,\ps)$ the unit sphere of
$\cC(X,\ps)$ and by $P_{X,\ps}$ the projection of $\cH$ onto
$\cC(X,\ps)$.
A closed subspace of $\cH$ is said to be {\em invariant} 
under $X$ iff it is invariant under all projections
$E^{X}(\De)$ for $\De\in\cB(\R)$.
Since  $\cC(X,\ps)$ is invariant under $X$, the projection $P_{X,\ps}$
commutes with $E^{X}(\De)$ for all $\De\in\cB(\R)$.
Then we obtain the following theorem.

\begin{Theorem}\label{th:PC-state2}
For any two observables $X$ and $Y$ on $\cH$ and any
state vector $\ps\in\cH$,
the following conditions are equivalent.

(i) $X$ and $Y$ are perfectly correlated in $\ps$.

(ii) $X$ and $Y$ are perfectly correlated in any 
$\ph\in\cC_{1}(X,\ps)$.

(iii) $E^{X}(\De)\ps=E^{Y}(\De)\ps$ for any $\De\in\cB(\R)$.

(iv) $f(X)\ps=f(Y)\ps$ for any $f\in B(\R)$.

(v) $f(X)P_{X,\ps}=f(Y)P_{X,\ps}$ for any $f\in B(\R)$.

(vi) $P_{X,\ps}=P_{Y,\ps}$ and $XP_{X,\ps}=YP_{Y,\ps}$.
\end{Theorem}
\begin{Proof}
Suppose (i) holds.
Let $\De\in\cB(\R)$.  Then, we have
\beqas
\|E^{X}(\De)\ps-E^{Y}(\De)\ps\|^{2}
&=&
\|E^{X}(\De)\ps\|^{2}
-\bracket{E^{X}(\De)\ps,E^{Y}(\De)\ps}\\
& &\mb{}-\bracket{E^{Y}(\De)\ps,E^{X}(\De)\ps}
+\|E^{Y}(\De)\ps\|^{2}
=0.
\eeqas
Thus, we have $E^{X}(\De)\ps=E^{Y}(\De)\ps$ for every $\De\in\cB(\R)$,
and the implication (i)$\Then$(iii) follows.
Suppose (iii) holds.
The set of Borel functions $f\in B(\R)$ satisfying $f(X)\ps=f(Y)\ps$ is
closed under the linear combination,  the uniform convergence, 
and includes all characteristic functions $\ch_{\De}$ for $\De\in\cB(\R)$, 
so that $f(X)\ps=f(Y)\ps$ holds for every $f\in B(\R)$.  Thus,
the implication (iii)$\Then$(iv) follows.
Suppose that condition (iv) holds. 
Then, we have
$f(X)g(X)\ps=f(Y)g(Y)\ps=f(Y)g(X)\ps$ for any $f,g\in B(\R)$. Since every
$\ph\in\cC(X,\ps)$ is a limit of vectors of the form $\ph=g(X)\ps$ 
for some $g\in B(\R)$, we have
$f(X)P_{X,\ps}=g(Y)P_{X,\ps}$. Thus, the implication (iv)$\Then$(v)
follows. 
Suppose that condition (v) holds.
The implication (v)$\Then$(iv) trivially holds,
and hence we have $\cC(X,\ps)=\cC(Y,\ps)$ and $P_{X,\ps}=P_{Y,\ps}$.
Letting $f=\ch_{\De}$ in condition $(v)$, we have 
$E^{X}(\De)P_{X,\ps}=E^{Y}(\De)P_{Y,\ps}$, and hence the spectral
measures of the self-adjoint operators $XP_{X,\ps}$ and $YP_{Y,\ps}$
are the same, so that they are identical.
Thus, the implication (v)$\Then$(vi) follows.
Suppose that condition (vi) holds.
Let $\ph\in\cC(X,\ps)$ and $\De,\Ga\in\cB(\R)$.
By the assumption we have $E^{X}(\Ga)P_{X,\ps}=E^{Y}(\Ga)P_{Y,\ps}$, 
so that we have $E^{X}(\Ga)\ph=E^{Y}(\Ga)\ph$,
and hence 
$$
\bracket{E^{X}(\De)\ph,E^{Y}(\Ga)\ph}=
\bracket{E^{X}(\De)\ph,E^{X}(\Ga)\ph}=
\bracket{\ph,E^{X}(\De\cap\Ga)\ph}.
$$
It follows that $X$ and $Y$ are perfectly correlated in $\ph$,
and hence the implication (vi)$\Then$(ii) follows.
Since the implication (ii)$\Then$(i) is obvious, the proof is completed.
\end{Proof}

It should be noticed that condition (vi) above does not imply the relation
$X\ps=Y\ps$, since $\ps$ may not be in the domain of $X$ or $Y$.  
However, for any rapidly decreasing $f$, i.e., $f\in\cS(\R)$,
we have $f(X)\ps$ is in the domains of $X$ and $Y$, and
that the self-adjoint extension of $XP_{X,\ps}-YP_{Y,\ps}$
coincides with the zero operator.  

For bounded $X$ and $Y$,
condition (vi) above is equivalent to that
$X\ph=Y\ph$ for all $\ph\in\cC_{1}(X,\ps)$, and the later condition
means that the observable $X-Y$ has the definite value zero in state
$\ph$, so that it is an interesting question to ask whether the relation
\beqa\label{eq:zero-difference}
X\ps=Y\ps
\eeqa
ensures that $X$ and $Y$ are perfectly correlated in $\ps$.  
If bounded observables $X$ and $Y$ commute, 
by multiplying $f(X)$ to the both sides we have $Xf(X)\ps=Yf(X)\ps$
for all $f\in B(\R)$ so that we have $XP_{X,\ps}=YP_{X,\ps}$,
and hence $X$ and $Y$ are perfectly correlated in $\ps$.
Busch, Heinonen, and Lahti \cite{BHL04} pointed out that
\Eq{zero-difference} does not ensure that $X$ and $Y$ are identically 
distributed in $\ps$.
Here, we shall show that 
even unitarily equivalent $X$ and $Y$ satisfying
\Eq{zero-difference} may fail to be perfectly correlated.
Let $X, Y$  and $\ps$ be two $4\times 4$ matrices and a 
$4$ dimensional column vector such that
\beqas
X=\left(
\begin{array}{cccc}
1 & 1 & 0 & 0\\
1 & 1& 0 & 0\\
0 & 0 & 1 & 1\\
0 & 0 & 1 & 0
\end{array}
\right),
\quad
Y=\left(
\begin{array}{cccc}
1 & 1 & 0 & 0\\
1 &0& 0 & 0\\
0 & 0 & 1 & 1\\
0 & 0 & 1 &1
\end{array}
\right),
\quad
\ps=
\left(
\begin{array}{c}
 1 \\
0 \\
 0\\
 0
\end{array}
\right).
\eeqas
Then, it is easy to see that $X$ and $Y$ are unitarily equivalent and 
satisfy \Eq{zero-difference}.
However, we have $\bracket{\ps|X^{3}|\ps}=4$ 
but $\bracket{\ps|Y^{3}|\ps}=3$.
Thus, the third moments of $X$ and $Y$ are different,  
so that the observables $X$ and $Y$ have different probability
distributions in $\ps$, and hence from Proposition \ref{th:equivalence-POVM}
they cannot be perfectly correlated.

Let $X$ be an observable on $\cH$ and $\rh$ a state on $\cH$.
The cyclic subspace of $\cH$ spanned by observable 
$X$ and state $\rh$ is the closed subspace $\cC(X,\rh)$ 
defined by
\beqa\label{eq:PCMS-1}
\cC(X,\rh)=\mb{the closure of }\{f(X)\ps\in\cH|\ 
f\in B(\R), \ \ps\in\ran(\rh)\}.
\eeqa
Then, it is easy to see the following relation 
\beqa\label{eq:PCMS-2}
\cC(X,\rh)=\mb{the closure of }\bigcup_{\ps\in\ran(\rh)}\cC(X,\ps).
\eeqa
In particular we have $\cC(X,\ketbra{\ps})=\cC(X,\ps)$
for any state vector $\ps\in\cH$.
Denote by $\cC_{1}(X,\rh)$ the unit sphere of $\cC(X,\rh)$ and
by $P_{X,\rh}$ the projection of $\cH$ onto $\cC(X,\rh)$.
Since $P_{X,\rh}=\bigvee_{\ps\in\ran(\rh)}P_{X,\ps}$, we have
$[P_{X,\rh},E^{X}(\De)]=0$ for all $\De\in\cB(\R)$.
Denote by $\cS(X,\rh)$ the space of states supported in $\cC(X,\rh)$,
i.e.,
\beqa\label{eq:PCMS-3}
\cS(X,\rh)=\{\si\in\cS(\cH)|\ \ran(\si)\subseteq\cC(X,\rh)\}.
\eeqa
It is easy to see that the following conditions are equivalent:
(i) $\si\in\cS(X,\rh)$. (ii) $P_{X,\rh}\si=\si$.
(iii)  $\si P_{X,\rh}=\si$. (iv) $P_{X,\rh}\si P_{X,\rh}=\si$.

Then, we obtain the following characterization of perfect correlation
in a mixed state.

\begin{Theorem}
\label{th:PCMS}
For any two observables $X$ and $Y$ on $\cH$ and any state
$\rh$ on $\cH$, the following conditions are equivalent.

(i) $X$ and $Y$ are perfectly correlated in $\rh$.

(ii) $X$ and $Y$ are perfectly correlated in any $\ps\in\ran(\rh)$.

(iii) $X$ and $Y$ are perfectly correlated in any $\ps\in\cC_{1}(X,\rh)$.

(iv) $X$ and $Y$ are perfectly correlated in any $\si\in\cS(X,\rh)$.

(v) $E^{X}(\De)\rh=E^{Y}(\De)\rh$ for any $\De\in\cB(\R)$.

(vi) $f(X)P_{X,\rh}=f(Y)P_{X,\rh}$.

(vii) $P_{X,\rh}=P_{Y,\rh}$ and $XP_{X,\rh}=YP_{Y,\rh}$.
\end{Theorem}
\begin{Proof}
Suppose (i) holds.
Let $\De\in\cB(\R)$.
From Proposition \ref{th:equivalence-POVM},
we have
\beqas
\lefteqn{
\|E^{X}(\De)\sqrt{\rh}-E^{Y}(\De)\sqrt{\rh}\|_{HS}^{2}
}\quad\\
&=&\Tr[E^{X}(\De)\rh]-\Tr[E^{X}(\De)E^{Y}(\De)\rh]
-\Tr[E^{Y}(\De)E^{X}(\De)\rh]+\Tr[E^{Y}(\De)\rh]=0,
\eeqas
where $\|\cdot\|_{HS}$ stands for the Hilbert-Schmidt norm. 
It follows that we have
$E^{X}(\De)\rh=E^{Y}(\De)\rh$, and hence the implication
(i)$\Then$(v) follows.
The implications (v)$\Then$(ii), (ii)$\Then$(iii), 
and (iii)$\Then$(vi) follow easily from the implications
(iii)$\Then$(i), (i)$\Then$(ii), and (ii)$\Then$(iv) in Theorem
\ref{th:PC-state2}, respectively. Assume condition (vi).  
It follows immediately that $f(X)\ps=f(Y)\ps$ for all $\ps\in\ran(\rh)$,
so that from \Eq{PCMS-1} we have $\cC(X,\rh)=\cC(Y,\rh)$
and $P_{X,\rh}=P_{Y,\rh}$.  Then, by assumption we have
$E^{X}(\De)P_{X,\rh}=E^{Y}(\De)P_{Y,\rh}$ for all $\De\in\cB(\R)$,
and hence we conclude that $XP_{X,\rh}$ equals $YP_{Y,\rh}$ 
since their spectral measures coincides.
Thus, the implication (vi)$\Then$(vii) follows.
Assume condition (vi).
Suppose $\si\in\cS(X,\rh)$.
Then $P_{X,\rh}\si=\si$, so that 
$E^{X}(\De)\si=E^{Y}(\De)\si$ and it is easy to see that
$X$ and $Y$  are perfectly correlated in $\si$,
and the implication (vi)$\Then$(iv) follows.
The implication (iv)$\Then$(i) trivially holds, and the proof is completed.
\end{Proof}

For observables with a complete orthonormal family of eigenvectors  (discrete
observables), we have the following important characterization of perfectly
correlating states.

\begin{Theorem}\label{th:SCECE}
Two discrete observables $X$ and $Y$ are perfectly correlated
in a vector state $\ps$ if and only if $\ps$ is a superposition
of common eigenstates of $X$ and $Y$ with common
eigenvalues.
\end{Theorem}
\begin{Proof}
Suppose that $X$ and $Y$ are perfectly correlated in a
state $\ps$.  Then, $\cC(X,\ps)$
is generated by eigenstates of $XP_{X,\ps}=YP_{X,\ps}$.
Thus, $\ps$ is a superposition of common eigenstates of 
$X$ and $Y$ with common eigenvalues.
Conversely, suppose that  $\ps$ is a superposition
of common eigenstates of $X$ and $Y$ with common
eigenvalues.
Then, the subspace $\cS$ generated by those eigenstates is
invariant under both $X$ and $Y$ and includes $\ps$.
Thus, $\cC(X,\ps)\subseteq\cS$, and $X=Y$ on $\cC(X,\ps)$,
and hence from Theorem \ref{th:PC-state2}, we conclude $X$  and $Y$
are perfectly correlated in $\ps$.
\end{Proof}

We say that two observables $X$ and $Y$ are {\em identically
distributed}  in a state $\rh$ iff $\Tr[E^{X}(\De)\rh]=\Tr[E^{Y}(\De)\rh]$
for all $\De\in\cB(\R)$.  Then, we have the following.

\begin{Theorem}\label{th:Born}
For any two observables $X$ and $Y$ on $\cH$ and any
state $\rh\in\cS(\cH)$, the following conditions are equivalent.

(i) $X$ and $Y$ are perfectly correlated in state $\rh$.

(ii) $X$ and $Y$ are identically distributed in any 
$\ps\in\cC_{1}(X,\rh)$.

(iii) $X$ and $Y$ are identically distributed in any state 
$\rh\in\cS(X,\rh)$.
\end{Theorem}
\begin{Proof}\sloppy
Suppose (i) holds.  Let $\De\in\cB(\R)$.
From Theorem \ref{th:PCMS}, we have 
$E^{X}(\De)\si=E^{Y}(\De)\si$ for any $\si\in\cS(X,\rh)$.
Thus, (iii) holds, and (i)$\Then$(iii) follows.
The implication (iii)$\Then$(ii) is obvious.
Suppose that (ii) holds.
Let $\ps\in\cC_{1}(X,\rh)$.
Let $\De,\Ga$ be disjoint Borel sets in $\cB(\R)$.
Then, $E^{X}(\De)\ps\in\cC(X,\rh)$, and hence
$$
\bracket{E^{X}(\De)\ps,E^{Y}(\Ga)E^{X}(\De)\ps}
=
\bracket{E^{X}(\De)\ps,E^{X}(\Ga)E^{X}(\De)\ps}
=0.
$$
Thus, by the Schwarz inequality we have
$$
|\bracket{E^{X}(\De)\ps,E^{Y}(\Ga)\ps}|^{2}
\le
\|E^{Y}(\Ga)E^{X}(\De)\ps\|^{2}
=
\bracket{E^{X}(\De)\ps,E^{Y}(\Ga)E^{X}(\De)\ps}
=
0.
$$
It follows that (i) holds.  Thus, the proof is completed.
\end{Proof}

It should also be noticed that even two identically distributed 
commuting observables $X$ and $Y$ may fail to satisfy \Eq{zero-difference}.
To see this, suppose that $\cH=\cK\otimes\cK$ for some Hilbert space $\cK$.
Let $X=A\otimes I$ and $Y=I\otimes A$ for some bounded operator $A$
on $\cK$ and $\ps=\ph\otimes\ph$ for some state vector $\ph\in\cK$.
Then, we have $\bracket{\ps|E^{X}(\De)|\ps}=\bracket{\ph|E^{A}(\De)|\ph}
=\bracket{\ps|E^{Y}(\De)|\ps}$ for all $\De\in\cB(\R)$, and hence
they are identically distributed.  However, we have
$(X-Y)\ps
=A\ph\otimes \ph-\ph\otimes A\ph$,
and hence \Eq{zero-difference} does not hold unless $\ph$ is an
eigenvector of $A$.

\section{Transitivity of perfect correlations}\label{se:5}

We denote by $\{X=Y\}$ the subspace spanned by all states 
$\ps\in\cH$ such that
$X$ and $Y$ are perfectly correlated in $\ps$, i.e.,
$$
\{X=Y\}=\{\ps\in\cH\mid \bracket{E^{X}(\De)\ps,E^{Y}(\Ga)\ps}
=0 \mb{ for all disjoint Borel sets $\De,\Ga$}\}.
$$
We shall call $\{X=Y\}$ the {\em perfectly correlative domain} for $X$ and $Y$.
Then, we have

\begin{Theorem}\label{th:largest}
The space $\{X=Y\}$ is the largest closed subspace $\cK$
of $\cH$ satisfying the following conditions.

(i) $\cK$ is invariant under 
$X$ and $Y$ for all $\De\in\cB(\R)$.

(ii) $E^{X}(\De)\ps=E^{Y}(\De)\ps$ for all $\De\in\cB(\R)$
and $\ps\in\cK$.
\end{Theorem}
\begin{Proof}
Assume $\ps\in\{X=Y\}$.
Then, we have $E^{X}(\De)E^{X}(\Ga)\ps=
E^{X}(\De\cap\Ga)\ps=E^{Y}(\De\cap\Ga)\ps
=E^{Y}(\De)E^{Y}(\Ga)\ps=E^{Y}(\De)E^{X}(\Ga)\ps$.
Thus, $\{X=Y\}$ is invariant under $X$,
and similarly under $Y$.  The space
$\{X=Y\}$ satisfies condition (ii) obviously from
Theorem \ref{th:PC-state2}.
Assume that $\cK$ satisfies conditions (i) and (ii).
Let $\ps\in\cK$.  Then, from  (ii) we have $\ps\in
\{X=Y\}$, and hence $\{X=Y\}$ is the largest.
\end{Proof}

From the above theorem, 
$\ps\in\{X=Y\}$ if and only if 
$\cC(X,\ps)\subseteq\{X=Y\}$.
The following theorem shows that the perfect correlation
in a given state is an equivalence
relation between observables.

\begin{Theorem}\label{th:transitivity}
For any observables $X,Y,Z$, we have $\{X=X\}=\cH$, $\{X=Y\}=\{Y=X\}$,
and $\{X=Y\}\cap\{Y=Z\}\subseteq\{X=Z\}$.
\end{Theorem}
\begin{Proof}
The relations $\{X=X\}=\cH$ and $\{X=Y\}=\{Y=X\}$ are obvious.
Let $\ps\in\{X=Y\}\cap\{Y=Z\}$ and $f\in B(\R)$.  
Then, we have $f(X)\ps=f(Y)\ps$ and $f(Y)\ps=f(Z)\ps$,
so that $f(X)\ps=f(Z)\ps$. Since $f$ is arbitrary, we have $\ps\in\{X=Z\}$.
Thus, we conclude $\{X=Y\}\cap\{Y=Z\}\subseteq\{X=Z\}$.
\end{Proof}

We denote by $\val{X=Y}$ the projection of $\cH$ onto
$\{X=Y\}$.  From Theorem \ref{th:largest} we have
\beq\label{eq:largest}
[E^{X}(\De)-E^{Y}(\De)]\val{X=Y}=0
\eeq
for all $\De\in\cB(\R)$.

\begin{Theorem}\label{th:PC-mixed-states}
For any two observables $X$ and $Y$ on $\cH$ and any state $\rh$
on $\cH$, the following conditions are equivalent.

(i) $X$ and $Y$ are perfectly correlated in $\rh$.

(ii) $\ran(\rh)\subseteq\{X=Y\}$ .

(iii) $\val{X=Y}\rh=\rh$.

(iv) $\rh\val{X=Y}=\rh$.
\end{Theorem}
\begin{Proof}
Suppose that $X$ and $Y$ are perfectly correlated in a state
$\rh$.
Let $\ps\in\ran(\rh)\setminus\{0\}$.
Then,  $\ps=\sqrt{\rh}\ph$ 
for some vector $\ph\in\cH$.
For any disjoint $\De,\Ga\in\cB(\R)$ 
we have 
\beqas
\bracket{E^{X}(\De)\ps,E^{Y}(\Ga)\ps}& =&
\|\ph\|^{2}
\bracket{\ph/\|\ph\|,
\sqrt{\rh}E^{X}(\De)E^{Y}(\Ga)\sqrt{\rh}\ph/\|\ph\|}\\
& \le&\|\ph\|^{2}\Tr[\sqrt{\rh}E^{X}(\De)E^{Y}(\Ga)\sqrt{\rh}]
=0.
\eeqas
Thus, $\ps\in\{X=Y\}$, so that $\ran(\rh)\subseteq\{X=Y\}$,
and the implication (i)$\Then$(ii) follows.
The implication (ii)$\Then$(iii) is obvious.
Suppose $\val{X=Y}\rh=\rh$.
From \Eq{largest}, we have
$E^{X}(\De)\rh=E^{Y}(\De)\rh$
and hence $X$ and $Y$ are perfectly correlated in $\rh$,
and the implication (iii)$\Then$(i) follows.
The equivalence (iii)$\Iff$(iv) follows immediately from taking
the adjoint of the both sides of relation (iii) or (iv).
\end{Proof}

For two observables $X,Y$ and a state $\rh$,
we denote by $X\equiv_{\rh} Y$ iff $X$ and $Y$ are perfectly
correlated in $\rh$.
The following theorem shows that the relation $\equiv_{\rh}$ 
is an equivalence relation between observables, and in particular
it is transitive.

\begin{Theorem}
For any observables $X,Y,Z$ and state $\rh$, we have
(i) $X\equiv_{\rh}X$, (ii) if $X\equiv_{\rh}Y$ then $Y\equiv_{\rh}X$,
and (iii) if $X\equiv_{\rh}Y$ and $Y\equiv_{\rh}Z$ then $X\equiv_{\rh}Z$.
\end{Theorem}
\begin{Proof}
From Theorems \ref{th:transitivity} and \ref{th:PC-mixed-states},
statements (i) and (ii) follow easily.
Suppose $X\equiv_{\rh}Y$ and $Y\equiv_{\rh}Z$.
Then, from Theorem \ref{th:PC-mixed-states}
we have $\ran(\rh)\subseteq \{X=Y\}$ 
and $\ran(\rh)\subseteq \{Y=Z\}$, 
and hence $\ran(\rh)\subseteq\{X=Y\}\cap\{Y=Z\}$.
From Theorem \ref{th:transitivity}, we have $\ran(\rh)\subseteq\{X=Z\}$,
so that we have shown $X\equiv_{\rh}Z$, and statement (iii) follows.
\end{Proof}

\section{Joint distributions}\label{se:6}
\subsection{Perfect correlations and joint probability distributions}

Let $X$ and $Y$ be two observables on $\cH$.
We say that $X$ and $Y$ {\em commute} on 
a closed subspace $\cK\subseteq\cH$ iff $\cK$ is
invariant under $X$ and $Y$ and 
$[E^{X}(\De),E^{Y}(\Ga)]\ps=0$
for all $\De,\Ga\in\cB(\R)$ and $\ps\in\cK$.
The {\em commutative domain}  of $X$
and $Y$ is defined 
to be the set $\com(X,Y)$ of those vectors $\ps\in\cH$ such that 
$
[E^{X}(\De),E^{Y}(\Ga)]\ps=0
$
for all $\De,\Ga\in\cB(\R)$.
It is clear that if $X$ and $Y$ commute on $\cK$ then
$\cK\subseteq\com(X,Y)$.  It can be easily seen that 
$\com(X,Y)$ is invariant under $X$ and $Y$; in fact,
if $\ps\in\com(X,Y)$, we have 
$E^{X}(\De_{1})E^{Y}(\De_{2})E^{X}(\De_{3})\ps
=
E^{X}(\De_{1})E^{X}(\De_{3})E^{Y}(\De_{2})\ps
=
E^{X}(\De_{1}\cap\De_{3})E^{Y}(\De_{2})\ps
=
E^{Y}(\De_{2})E^{X}(\De_{1}\cap\De_{3})\ps
=
E^{Y}(\De_{2})E^{X}(\De_{1})E^{X}(\De_{3})\ps,
$
so that $E^{X}(\De_{3})\ps\in\com(X,Y)$.
Thus, $\com(X,Y)$ is the largest closed subspace 
on which $X$ and $Y$ commute; see Ylinen \cite{Yli85}. 
Let $C_{X,Y}$ denote the projection of $\cH$ 
onto $\com(X,Y)$.
Then, we have
\beqa\label{eq:com}
[E^{X}(\De),E^{Y}(\Ga)]C_{X,Y}=0
\eeqa
for all $\De,\Ga\in\cB(\R)$.  
The following theorem generalizes Yilnen's theorem \cite{Yli85} on
characterization of pure states in  $\com(X,Y)$ to mixed states.

\begin{Theorem}\label{th:JPD}
For any state $\rh$, the following conditions
are equivalent.

(i) $C_{X,Y}\rh=\rh$.

(ii) There is a spectral measure $E$ on $\cB(\R^{2})$ such that
$E(\De\times\Ga)\rh=E^{X}(\De)\And E^{Y}(\Ga)\rh$ for all $\De,\Ga\in
\cB(\R)$.

(iii) The function $\De\times\Ga\mapsto\Tr[E^{X}(\De)\And
E^{Y}(\Ga)\rh]$ on $\cB(\R)\times\cB(\R)$ extends 
to a probability measure
on
$\cB(\R^{2})$.

(iv) $E^{X}(\De)E^{Y}(\Ga)\rh
=E^{Y}(\Ga)E^{X}(\De)\rh$
for all $\De,\Ga\in\cB(\R)$.

\end{Theorem}
\begin{Proof}
Since $XC_{X,Y}$ and $YC_{X,Y}$ are commuting
self-adjoint operators, there is another self-adjoint operator
$Z$ and two real-valued Borel functions $f,g$
such that
$XC_{X,Y}=f(Z)$ and $YC_{X,Y}=g(Z)$ \cite{vN32}. 
Let $E$ be the spectral measure on $\cB(\R^{2})$ 
defined by $E(\De\times\Ga)
=E^{Z}(f^{-1}(\De)\cap g^{-1}(\Ga))$
for all $\De,\Ga\in\cB(\R)$.
Let $\De,\Ga\in\cB(\R)$.
We have $E(\De\times\Ga)C_{X,Y}
=E^{Z}(f^{-1}(\De))E^{Z}(g^{-1}(\Ga))C_{X,Y}
=E^{X}(\De)E^{Y}(\Ga)C_{X,Y}
=E^{X}(\De)\And E^{Y}(\Ga)C_{X,Y}.$
Thus, it is easy to see that the implication  (i)$\Then$(ii) follows.
The implication (ii)$\Then$(iii) follows obviously.
Assume condition (iii).
Let $\mu$ be the probability measure on $\cB(\R^{2})$ such that
$\mu(\De\times\Ga)=\Tr[E^{X}(\De)\And E^{Y}(\Ga)\rh]$.
Let $P=E^{Y}(\Ga)-E^{X}(\De)\And E^{Y}(\Ga)
-E^{X}(\R\setminus\De)\And E^{Y}(\Ga)$.
Then, $P$ is a projection and $E^{X}(\De)P=
E^{X}(\De)E^{Y}(\Ga)-E^{X}(\De)\And E^{Y}(\Ga)$.
By the countable additivity
of $\mu$, we have
$
\Tr[(P\sqrt{\rh})^{\da}(P\sqrt{\rh})]=\Tr[P\rh]=
\mu(\R\times\Ga)-\mu(\De\times\Ga)-
\mu((\R\setminus\De)\times\Ga)
=0.
$
Thus, we have 
$P\sqrt{\rh}=0$ so that $E^{X}(\De)P\rh=0$, and hence we have
$E^{X}(\De)E^{Y}(\Ga)\rh=E^{X}(\De)\And E^{Y}(\Ga)\rh$.
By symmetry, we also obtain 
$E^{Y}(\Ga)E^{X}(\De)\rh=E^{X}(\De)\And E^{Y}(\Ga)\rh$.
Thus, the implication (iii)$\Then$(iv) follows.
Assume condition (iv).
Then, we have $\rh\ps\in\com(X,Y)$ for all $\ps\in\cH$.
Thus, $C_{X,Y}\rh\ps=\rh\ps$ for all $\ps\in\cH$,
and hence the implication (iv)$\Then$(i) follows.
\end{Proof}

Observables $X$ and $Y$ are said to be {\em compatible} 
in a state $\rh$ iff $C_{X,Y}\rh=\rh$,
and they are said to have the {\em joint probability distribution}
in $\rh$ iff there is a probability measure 
$\mu^{X,Y}_{\rh}$ on $\cB(\R^{2})$ 
satisfying 
\beqa
\mu^{X,Y}_{\rh}(\De\times\Ga)=
\Tr[E^{X}(\De)\And E^{Y}(\Ga)\rh]
=
\Tr[E^{Y}(\Ga)E^{X}(\De)\rh]
=
\Tr[E^{X}(\De)E^{Y}(\Ga)\rh]
\eeqa
for all $\De,\Ga\in\cB(\R)$.
Theorem \ref{th:JPD} shows that $X$ and $Y$ have the joint 
probability distribution in $\rh$ if and only if they are
compatible in $\rh$.

Two observables $X$ and $Y$ are called {\em jointly measurable}
in a state $\rh$ iff they have the joint probability distribution
$\mu^{X,Y}_{\rh}$ and satisfy the following relations
\beqa
\mu^{X,Y}_{\rh}(\De\times\Ga)
&=&\Tr[E^{X}(\De)E^{Y}(\Ga)E^{X}(\De)\rh],
\label{eq:JO1}\\
\mu^{X,Y}_{\rh}(\De\times\Ga)
&=&\Tr[E^{Y}(\Ga)E^{X}(\De)E^{Y}(\Ga)\rh]
\label{eq:JO2}
\eeqa
for any $\De,\Ga\in\cB(\R)$.
The above relations ensure that the theoretical joint probability of
the event ``$X\in\De$ and $Y\in\Ga$'' is obtained as the joint probability
of outcomes of the successive projective measurements of projections
$E^{X}(\De)$ and $E^{Y}(\Ga)$ irrespective of the order of the 
measurements \cite{01OD}.  Moreover,
for discrete observables $X$ and $Y$, the above relation ensures
that the the joint probability distribution of the outcomes of the successive
projective measurements of observables $X$ and $Y$ coincides with
the joint probability distribution $\mu_{X,Y}$ 
irrespective of the order of the measurements.

\begin{Theorem}\label{th:JM}
Every pair of  observables $X$ and $Y$ compatible
in a state $\rh$ is jointly measurable in the state $\rh$.
\end{Theorem}
\begin{Proof}
The assertion follows immediately from Theorem \ref{th:JPD}.
\end{Proof}

Denote by $\bD$ the diagonal set in $\R^{2}$, i.e.,
$\bD=\{(x,y)\in\R^{2}\mid x=y\}$.

\begin{Theorem}\label{th:PC-JPD}
Two observables $X$ and $Y$ are
perfectly correlated in a sate $\rh$ 
if and only if $X$ and $Y$ are compatible in $\rh$
and the joint probability distribution is concentrated in
the diagonal set, i.e., 
$\mu^{X,Y}_{\rh}(\R\setminus\bD)=0$.
\end{Theorem}
\begin{Proof}
If $x\not=y$, there is a rational number $q$ such that
$x,y>q$ or $x,y<q$, and hence it is easy to see that
\beqa
\R\setminus\bD=\bigcup_{q\in\Q}(-\infty,q)\times(q,\infty)
\cup\bigcup_{q\in\Q}(q,\infty)\times(-\infty,q),
\eeqa
where $\Q$ stands for the set of rational numbers.
Suppose that $X$ and $Y$ are perfectly correlated
in $\rh$.
Then, we have
$
E^{X}(\De)E^{Y}(\Ga)\rh
=E^{X}(\De\cap\Ga)\rh
=E^{Y}(\De\cap\Ga)\rh
=E^{Y}(\Ga)E^{Y}(\De)\rh
=E^{Y}(\Ga)E^{X}(\De)\rh,
$
and hence $X$ and $Y$ are compatible in $\rh$.
Accordingly, 
the joint probability distribution satisfies
$\mu^{X,Y}_{\rh}((-\infty,q)\times(q,\infty))
=
\mu^{X,Y}_{\rh}((q,\infty)\times(-\infty,q))
=0,
$
so that $\mu^{X,Y}_{\rh}(\R\setminus\bD)=0$.
Conversely, suppose that $X$ and $Y$ are compatible in $\rh$
and $\mu^{X,Y}_{\rh}(\R\setminus\bD)=0$.
Let $\De,\Ga\in\cB(\R)$.
In general, we have $(\De\times\Ga)\cap\bD=[\R\times(\De\cap\Ga)]\cap\bD$.
Thus, if $\De\cap\Ga=\emptyset$, we have
\begin{eqnarray}
\Tr[E^{X}(\De)E^{Y}(\Ga)\rh] 
= \mu^{X,Y}_{\rh}((\De\times\Ga)\cap\bD)
= \mu^{X,Y}_{\rh}([\R\times(\De\cap\Ga)]\cap\bD)
= 0,
\end{eqnarray}
so that $X$ and $Y$ are perfectly correlated in $\rh$.
\end{Proof}

Let $\ep>0$.
Let $\cdots <\mu_{-1}<\mu_{0}<\mu_{1}<\cdots$ be a partition of
the real line $\R$ such that $\mu_{j+1}-\mu_{j}<\ep$ for all $j$.
Let $X_{\ep}$ and $Y_{\ep}$ be $\ep$ approximations of 
observables $X$ and $Y$ defined by
$X_{\ep}=\sum_{j\in\bZ}\la_{j}E^{X}(\De_{j})$
and 
$Y_{\ep}=\sum_{j\in\bZ}\la_{j}E^{Y}(\De_{j})$
where $\De_{j}=[\mu_{j},\mu_{j})$ and $\la_{j}\in\De_{j}$.
If $X$ and $Y$ are discrete observables, there are $\ep$
approximations $X_{\ep}$ and $Y_{\ep}$ such that
$X=X_{\ep}$ and $Y=Y_{\ep}$.
From Theorems \ref{th:JM} and \ref{th:PC-JPD} we conclude that
two observables $X$ and $Y$ perfectly correlated in a state $\rh$ 
have the joint probability distribution concentrated in the diagonal set 
and that each instance of the successive projective measurements of
any $\ep$ approximations $X_{\ep}$ and $Y_{\ep}$ gives the same output
irrespective of the order of the measurements for any $\ep>0$.

\subsection{Perfect correlations and quasi-probability distributions}

In Ref.~\cite{Urb61}, Urbanik introduced the following formulation
for the quasi-joint probability distribution for any pair of observables,
generalizing the quasi-joint probability distribution of the position 
and the momentum first studied by Wigner \cite{Wig32} and Moyal \cite{Moy49}.
Let $\mu$ be a probability measure on $\cB(\R^{2})$.
To any pair of real numbers $x,y$ there corresponds the family of
lines $S^{a,b}_{t}$ given by the equation $ax+by=t$, where $t\in\R$.
Letting for every Borel subset $\De\subseteq\R$
\beqa
\mu_{a,b}(\De)=\mu(\bigcup_{t\in\De}S^{a,b}_{t})
\eeqa
we obtain a probability measure on $\R$.
It is well-know that $\mu$  is determined uniquely by the family of
probability measures $\mu_{a,b}$.
We suppose that for all pair of real numbers $a,b$ the linear combinations
$aX+bY$ are self-adjoint operators on $\cH$.  Consequently, for every pair
$a,b\in\R$ and every state vector $\ps$ the probability distribution
of $aX+bY$ is defined by
\beqa
\mu^{aX+bY}_{\ps}(\De)=\bracket{\ps,E^{aX+bY}(\De)\ps}
\eeqa
for all $\De\in\cB(\R)$.
Given a state vector $\ps$, a probability measure $\mu$ on $\R^{2}$ is
said to be {\em the joint probability distribution} of observables $X$ and $Y$,
iff  $\mu_{a,b}$ is equal to $\mu^{aX+bY}_{\psi}$.  
The joint probability
distribution so defined is uniquely determined, provided it exists.  
We shall denote by $\nu^{X,Y}_{\ps}$ the joint probability distribution of $X,Y$
in $\ps$.
We also denote by $\Ph_{a,b}^{\ps}$ the characteristic function of the
probability distribution $\mu^{aX+bY}_{\ps}$:
\beqas
\Ph_{a,b}^{\ps}(t)=\bracket{\ps,e^{it(aX+bY)}\ps}
\eeqas
for all $t\in\R$.  Then, from Bochner's theorem
it is easy to see that observables $X$ and $Y$ have
the joint probability distribution in a vector state $\ps\in\cH$ if and only if
the function $\Ph_{t,s}^{\ps}(1)$ of two variable $t,s$ is a continuous
positive definite function on $\R^{2}$.

\begin{Theorem}
For any observables $X,Y$ and any state vector $\ps$, the following
conditions are equivalent.

(i) $X$ and $Y$ are perfectly correlated in $\ps$.

(ii) $\Ph_{t,s}^{\ph}(1)=\Ph_{t+s,0}^{\ph}(1)$ for any $t,s\in\R$ and
$\ph\in\cC_{1}(X,\ps)$

(iii) $\Ph_{t,0}^{\ph}(1)=\Ph_{0,t}^{\ph}(1)$ for any $t\in\R$ and
$\ph\in\cC_{1}(X,\ps)$.
\end{Theorem}
\begin{Proof}
Assume (i) holds.
Then, we have $XP_{X,\ps}=YP_{X,\ps}$, so that
$e^{i(tX+sY)}P_{X,\ps}=e^{itX}e^{isY}P_{X,\ps}=e^{itX}e^{isX}P_{X,\ps}
=e^{i(t+s)X}P_{X,\ps}$.  
Let $\ph\in\cC_{1}(X,\ps)$.
Then, we have
 $\Ph_{t,s}^{\ph}(1)=\bracket{\ph,e^{i(tX+sY)}P_{X,\ps}\ph}
=\bracket{\ph,e^{i(t+s)X}\ph}$,
and hence the implication (i)$\Then$(ii) follows.
The implication (ii)$\Then$(iii) is obvious.
Assume (iii) holds.
Then, we have
\beqas
\bracket{\ph,e^{itX}\ph}=\bracket{\ph,e^{itY}\ph},
\eeqas
for all $t\in\R$.  It follows that $X$ and $Y$ are identically distributed in $\ph$.
Since $\ph\in\cC_{1}(X,\ps)$ is arbitrary, the implication (iii)$\Then$(i)
follows from Theorem \ref{th:Born}.
\end{Proof}

Our approach is more coherent with the following definition of
``characteristic functions''.
We define a function $\Ps_{a,b}^{\ps}$ on $\R$ by
\beqas
\Ps_{a,b}^{\ps}(t)=\bracket{e^{-itaX}\ps,e^{itbY}\ps}
\eeqas
for all $t\in\R$.  It is easy to see that $\Ps_{a,0}(t)=\Ph_{a,0}(t)$
and $\Ps_{0,a}(t)=\Ph_{0,a}(t)$ for all $a,b,t\in\R$.
Then, we have

\begin{Theorem}
For any observables $X,Y$ and any state vector $\ps$, the following
conditions are equivalent.

(i) $X$ and $Y$ are perfectly correlated in $\ps$.

(ii) $\Ps_{t,s}^{\ps}(1)=\Ps_{t+s,0}^{\ps}(1)$ for any $t,s\in\R$.
\end{Theorem}
\begin{Proof}
Suppose (i) holds.
From Theorem \ref{th:PC-state2} we have
$e^{isY}\ps=e^{isX}\ps$, and hence
\beqas
\Ps_{t,s}^{\ps}(1)
=\bracket{e^{-itX}\ps,e^{isY}\ps}
=\bracket{e^{-itX}\ps,e^{isX}\ps}
=\Ph_{t+s,0}^{\ps}(1)
\eeqas
for any $t,s\in\R$.
Suppose (ii) holds. 
We have
\beqas
\|e^{itX}\ps-e^{itY}\ps\|^{2}
=
2-2\Re\bracket{e^{itX}\ps,e^{itY}\ps}
=
2-2\Re\Ps_{-t,t}(1)
=0
\eeqas
Thus, we have $e^{itX}\ps=e^{itY}\ps$ for any $t\in\R$.
Since the von Neumann algebra generated by all $e^{itX}$
with $t\in\R$ coincides with that of all $f(X)$ with $f\in B(\R)$,
the set of Borel functions $f$ satisfying $f(X)\ps=f(Y)\ps$
includes $B(\R)$, and the implication (ii)$\Then$(i) follows.
\end{Proof}

\section{Perfect correlations and entanglement}
\label{se:7}

\subsection{Bipartite perfect correlations}
The notion of perfect correlation in quantum theory was 
discussed first by von Neumann \cite{vN32} to establish a quantum
mechanical description of a process of measurement
and is closely related to the notion of entanglement recently
discussed quite actively in the field of quantum information \cite{NC00}.
In what follows we shall discuss some examples in these fields.

Let $\cK_{1}$ and $\cK_{2}$ be two Hilbert spaces and suppose 
$\cH=\cK_{1}\otimes\cK_{2}$.
Every state vector $\ps$ has two orthonormal sequences $\{\ph_{j}\}$
and $\{\xi_{j}\}$ such that
\beqa\label{eq:entangled}
\ps=\sum_{j}\sqrt{p_{j}}\ph_{j}\otimes\xi_{j},
\eeqa
where $p_{j}>0$ and $\sum_{j} p_{j}=1$ \cite{vN32}.
The above decomposition is called the {\em Schmidt decomposition}
of $\ps$.
Then, the amount of entanglement \cite{NC00} of $\ps$ is defined by 
\beq
E(\ps)=-\sum_{j}p_{j}\log p_{j}.
\eeq
Let $\rh_{1}=\Tr_{2}\ketbra{\ps}$ and $\rh_{2}=\Tr_{1}\ketbra{\ps}$,
where $\Tr_{l}$ stands for the partial trace over $\cK_{l}$ for $l=1,2$.
Then, $E(\ps)=S(\rh_{1})=S(\rh_{2})$, where $S$ stands for the von Neumann
entropy, i.e., $S(\rh_{l})=-\Tr[\rh_{l}\log\rh_{l}]$ \cite{vN32}.
Let $X$ and $Y$ be observables on $\cH$
defined by $X=\sum_{j}\la_{j}\ketbra{\ph_{j}}$ 
and $Y=\sum_{j}\la_{j}\ketbra{\xi_{j}}$ 
with nondegenerate eigenvalues $\{\la_{j}\}$.
Then, we have $\bracket{E^{X\otimes I}(\{\la_{j}\})\ps,
E^{I\otimes Y}(\{\la_{k}\})\ps}=\sqrt{p_{j}p_{k}}
\bracket{\ph_{j}\otimes\xi_{j},\ph_{k}\otimes\xi_{k}}
=\de_{j,k}p_{j}$,
and hence we can conclude that $X\otimes I$ and $I\otimes Y$ are perfectly
correlated in $\ps$.  

\begin{Theorem}
Suppose $\cH=\cK_{1}\otimes\cK_{2}$
with $\dim(\cH)<\infty$. 
Let $\ps$ be a state on $\cH$.
For any two observables $X$ on 
$\cK_{1}$ and $Y$ on $\cK_{2}$,
the observables $X\otimes I$ and $I\otimes Y$ are 
perfectly correlated in $\ps$ if and only if
there is a pair of orthonormal basis $\{\ph_{j}\}$ of $\cK_{1}$
and $\{\xi_{j}\}$ of $\cK_{2}$ 
and a sequence of nonzero real numbers $\la_{1},\ldots,\la_{n}$
such that
$\ps$ has the Schmidt decomposition
$\ps=\sum_{j=1}^{n}\sqrt{p_{j}}\ph_{j}\otimes\xi_{j}$ 
with $p_{j}>0$ for all $j=1,\ldots,n$,
and that
$X\ph_{j}=\la_{j}\ph_{j}$ and 
$Y\xi_{j}=\la_{j}\xi_{j}$
for all $j=1,\ldots,n$.
\end{Theorem}
\begin{Proof}
Suppose that $X\otimes I$ and $I\otimes Y$ are 
perfectly correlated in $\ps$.
Then, by Theorem \ref{th:SCECE} the state $\ps$ is a superposition
of common eigenstates with common 
eigenvalues $\mu_{1},\ldots,\mu_{m}$ of $X$ and $Y$.
Let $\ps_{k}
=[E^{X}(\{\mu_{k}\})\otimes E^{Y}(\{\mu_{k}\})]\ps/\sqrt{q_{k}}$,
where $q_{k}=\|\Ps_{k}\|^{2}$
for all $k=1,\ldots,m$.
We have
$\ps=\sum_{k=1}^{m}\sqrt{q_{k}}\ps_{k}$ with $q_{k}>0$ and
$\sum_{k=1}^{m}q_{k}=1$.
Let $\ps_{k}
=\sum_{l=1}^{s(k)}\sqrt{r^{(k)}_{l}}\ph^{(k)}_{l}\otimes\xi^{(k)}_{l}$ 
be a Schmidt decomposition of $\ps_{k}$.
Then, 
$(X\otimes I)\ps_{k}
=\sum_{l=1}^{s(k)}
\sqrt{r^{(k)}_{l}}X\ph^{(k)}_{l}\otimes\xi^{(k)}_{l}$
and 
$(X\otimes I)\ps_{k}
=\sum_{l=1}^{s(k)}
\sqrt{r^{(k)}_{l}}\mu_{k}\ph^{(k)}_{l}\otimes\xi^{(k)}_{l}$.
Since
$\xi^{(k)}_{1},\ldots,\xi^{(k)}_{s(k)}$ are linearly independent, we have
$X\ph^{(k)}_{l}=\mu_{k}\ph^{(k)}_{l}$, and similarly we have 
$Y\xi^{(k)}_{l}=\mu_{k}\xi^{(k)}_{l}$.
Let $\la_{j}=\mu_{k}$ if $\sum_{l=1}^{k-1}s(l)<j\le\sum_{l=1}^{k}s(l)$,
let $\ph_{j}=\ph^{(k)}_{l}$, $\xi_{j}=\xi^{(k)}_{l}$,
$\sqrt{p_{j}}=\sqrt{q_{k}r^{(k)}_{l}}$ if $j=l+\sum_{l=1}^{k-1}s(l)$,
and let $n=\sum_{k=1}^{m}s(k)$.
Then, we have a Schmidt decomposition
$\ps=\sum_{j=1}^{n}\sqrt{p_{j}}\ph_{j}\otimes\xi_{j}$ 
with the desired properties.
The converse part is obvious from the discussion preceding the
present theorem, and the proof is completed.
\end{Proof}

\subsection{Nonlocality without inequality}

Let us consider the case where $\cH=\cK_{1}\otimes\cK_{2}$
and $\cK_{j}\cong\C^{2}$ for $j=1,2$.
Let $U,D$ be two observables on $\C^{2}$ having eigenvalues 
$1$ and $0$.
Let $U_{1}=U\otimes I$, $D_{1}=D\otimes I$,
$U_{2}=I\otimes U$, and $D_{1}=I\otimes D$.
Hardy \cite{Har93} showed that any state vector $\ps\in\cH$
shows nonlocality if it satisfies 
\beqa
P_{\ps}(U_{1}=0,U_{2}=1)&=&0,\label{eq:040426a}\\
P_{\ps}(U_{1}=1,D_{2}=0)&=&0,\label{eq:040426b}\\
P_{\ps}(D_{1}=1,U_{2}=0)&=&0,\label{eq:040426c}\\
P_{\ps}(D_{1}=1,D_{2}=0)&>&0,\label{eq:040426d}
\eeqa
where $P_{\ps}(A=a,B=b)=\bracket{E^{A}(\{a\})\ps,E^{B}(\{b\})\ps}$
for $A=U_{1}, D_{1}$, and $B=U_{2},D_{2}$, and $a,b=0,1$,
and showed that actually we can find such observables $U$ and $D$
for any state $\ps$ unless $\ps$ is a product state {\em or a maximally
entangled state}.
This failure of Hardy's nonlocality proof for the class of maximally 
entangled states has been explained by Cereceda \cite{Cer99} as follows:
the perfect correlation for pairs $(U_{1},U_{2})$, $(U_{1},D_{2})$,
and $(D_{1},U_{2})$  necessarily entails perfect correlation for the pair
$(D_{1},D_{2})$.
Now, we shall show that Cereceda's argument can be considerably simplified
by appealing to the general  property of the transitivity of perfect correlations.

Let $\ps$ be a general state vector in $\cH$.
Then, we have a Schmidt decomposition of $\ps$ such that
\beqa
\ps=\sqrt{p_{1}}\xi_{1}\otimes\et_{1}+\sqrt{p_{2}}\xi_{2}\otimes\et_{2}.
\eeqa
The numbers $0\le p_{2}\le p_{1}$ are uniquely determined
with $p_{1}+p_{2}=1$, and if $p_{1}\not=1/2,1$,
the vectors $\xi_{1}\otimes\et_{1}$ and $\xi_{2}\otimes\et_{2}$
are uniquely determined up to constant factors.
The essential part of Hardy's proof of nonlocality is that
if $1/2<p_{1}<1$, we can always find observables $U$
and $D$ such that $U\not=D$ while they satisfy 
Eqs.~\eq{040426a}--\eq{040426d}.
Now, suppose that $\ps$ is maximally entangled, i.e., $p_{1}=1/2$.
We shall show that \Eq{040426a} leads to  
$P_{\ps}(U_{1}=1,U_{2}=0)=0$.
Let $\{\xi_{0},\xi_{1}\}$ be an orthonormal basis such that
$U=\ketbra{\xi_{1}}$.
Expanding $\ps$ in the basis $\{\xi_{j}\otimes\xi_{k}\}_{j,k=0,1}$,
we have $\ps=\sum_{j,k}c_{jk}\xi_{j}\otimes\xi_{k}$.
Then, we have $P_{\ps}(U_{1}=j,U_{2}=k)=|c_{jk}|^{2}$ for all $j,k=0,1$.
From \Eq{040426a}, we have $c_{01}=0$, and hence we have
\beqas
\rh_{1}
&=&\Tr_{2}\ketbra{\ps}\\
&=&|c_{00}|^{2}\ketbra{\xi_{0}}
+c_{00}c_{10}^{*}\ket{\xi_{0}}\bra{\xi_{1}}
+c_{10}c_{00}^{*}\ket{\xi_{1}}\bra{\xi_{0}}
+(|c_{10}|^{2}+|c_{11}|^{2})\ketbra{\xi_{1}}.
\eeqas
Since $S(\rh_{1})=\log 2$, we have $\rh_{1}=I_{1}/2$,
and hence $|c_{00}|^{2}=1/2$ and $c_{10}c_{00}^{*}=0$,
so that we have $P_{\ps}(U_{1}=1,U_{2}=0)=|c_{10}|^{2}=0$.
Thus, \Eq{040426a} leads to $U_{1}\equiv_{\ps}U_{2}$.
Similarly, \Eq{040426b} leads to $U_{1}\equiv_{\ps}D_{2}$, and
\Eq{040426c} leads to $D_{1}\equiv_{\ps}U_{2}$.
Thus, by the transitivity of perfect correlation, we conclude 
$D_{1}\equiv_{\ps}D_{2}$, or $P_{\ps}(D_{1}=j,D_{2}=k)=0$
if $j\not=k$, and this contradicts \Eq{040426d}.

\section{Characterizations of perfectly correlated POVMs}
\label{se:4}
\subsection{Joint dilations of POVMs}

For any POVM $\Pi$, there is a triple
$(\cK,\xi,L)$,  called a {\em Naimark-Holevo dilation} of $\Pi$, consisting of a
separable Hilbert space
$\cK$, a state vector $\xi\in\cK$, and an observable $L$ on $\cH\otimes\cK$
satisfying
\beqa
\bracket{\ps',\Pi(\De)\ps}=
\bracket{\ps'\otimes\xi,E^{L}(\De)(\ps\otimes\xi)}
\eeqa
for any $\ps,\ps'\in\cH$ and $\De\in\cB(\R)$ \cite{Hol73SQ}.
We now extends the above notion to any pair of POVMs.
A {\em joint dilation} of POVMs $\Pi_{1},\Pi_{2}$ is a quadruple $(\cK,\xi,X,Y)$
consisting of a separable Hilbert space $\cK$, a
state vector $\xi\in\cK$, and observables $X,Y$ on
$\cH\otimes\cK$,  satisfying
\beqa\label{eq:joint_dilation}
\bracket{\Pi_{1}(\De)\ps',\Pi_{2}(\Ga)\ps}
=
\bracket{E^{X}(\De)(\ps'\otimes\xi),E^{Y}(\Ga)(\ps\otimes\xi)}
\eeqa
for all $\De,\Ga\in\cB(\R)$ and $\ps,\ps'\in\cH$.
In this case, we have
\beqa\label{eq:joint_dilation2}
\Tr[\Pi_{1}(\De)\Pi_{2}(\Ga)\rh]
=
\Tr[E^{X}(\De)E^{Y}(\Ga)(\rh\otimes\ketbra{\xi})]
\eeqa
for all $\De,\Ga\in\cB(\R)$ and $\rh\in\cH$.
The existence of the joint dilations is given in the following.

\begin{Theorem}\label{th:joint_dilation}
Any pair of POVMs has a joint dilation of them.
\end{Theorem}
\begin{Proof}
Let $\Pi_{1},\Pi_{2}$ be a pair of POVMs.
Let $(\cK_{j},\xi_{j},L_{j})$ be a Naimark-Holevo dilation
of $\Pi_{j}$ for $j=1,2$.
Let $\ph_{1},\ph_{2},\ldots$ be an arbitrary orthonormal basis of $\cH$.
Let $\et^{(j)}_{1},\et^{(j)}_{2},\ldots$ be an orthonormal basis of $\cK_{j}$ 
such that $\et^{(j)}_{1}=\xi_{j}$ for $j=1,2$.
Then, by repeated uses of the Parceval identity, for any $\ps,\ps'\in\cH$
and $\De,\Ga\in\cB(\R)$
we have
\beqas
\lefteqn{\bracket{\Pi_{1}(\De)\ps',\Pi_{2}(\Ga)\ps}}\quad\\
&=&
\sum_{k}\bracket{\Pi_{1}(\De)\ps',\ph_{k}}\bracket{\ph_{k},\Pi_{2}(\Ga)\ps}\\
&=&
\sum_{k}\bracket{E^{L_{1}}(\De)(\ps'\otimes\xi_{1}),\ph_{k}\otimes\xi_{1}}
\bracket{\ph_{k}\otimes\xi_{2},E^{L_{2}}(\Ga)(\ps\otimes\xi_{2})}\\
&=&
\sum_{k}\bracket{
(E^{L_{1}}(\De)\otimes I_{2})(\ps'\otimes\xi_{1}\otimes\xi_{2}),
\ph_{k}\otimes\xi_{1}\otimes\xi_{2}}\times\\
& &
\quad\times\bracket{
\ph_{k}\otimes\xi_{1}\otimes\xi_{2},
(E^{L_{2}}(\Ga)\otimes I_{1})(\ps\otimes\xi_{1}\otimes\xi_{2})}\\
&=&
\sum_{k,l,m}\bracket{
(E^{L_{1}}(\De)\otimes I_{2})(\ps'\otimes\xi_{1}\otimes\xi_{2}),
\ph_{k}\otimes\et_{l}^{(1)}\otimes\et_{m}^{(2)}}\times\\
& &
\quad\times\bracket{
\ph_{k}\otimes\et_{l}^{(1)}\otimes\et_{m}^{(2)},
(E^{L_{2}}(\Ga)\otimes I_{1})(\ps\otimes\xi_{1}\otimes\xi_{2})}\\
&=&
\bracket{
(E^{L_{1}}(\De)\otimes I_{2})(\ps'\otimes\xi_{1}\otimes\xi_{2}),
(E^{L_{2}}(\Ga)\otimes I_{1})(\ps\otimes\xi_{1}\otimes\xi_{2})},
\eeqas
whee $I_{j}$ is the identity operator on $\cK_{j}$.
Thus, we have a joint dilation $(\cK_{1}\otimes \cK_{2},
\xi_{1}\otimes\xi_{2},L_{1}\otimes I_{2},L_{2}\otimes
I_{1})$.
\end{Proof}

Using joint dilations, perfect correlations between POVMs are
reduced to those between observables.

\begin{Theorem}\label{th:distance-POVM}
For any joint dilation $(\cK,\xi,X,Y)$ of
a pair of POVMs $\Pi_{1},\Pi_{2}$,
the POVMs $\Pi_{1}$ and $\Pi_{2}$ are
perfectly correlated in a state $\rh\in\cS(\cH)$
if and only if $X$ and $Y$ are perfectly correlated in
$\rh\otimes\ketbra{\xi}$.  In this case, we have
\beqa\label{eq:PCPOVM}
\Pi_{1}(f)\rh=\Pi_{2}(f)\rh
\eeqa
for any $f\in B(\R)$.
\end{Theorem}
\begin{Proof}
Let $(\cK,\xi,X,Y)$ be a joint dilation of $\Pi_{1}$ and $\Pi_{2}$.
Then, from \Eq{joint_dilation2}
it is easy to see that $\Pi_{1}$ and $\Pi_{2}$ are perfectly
correlated in $\rh$ if and only if so are $X$ and $Y$ in
$\rh\otimes\ketbra{\xi}$. In this case, from Theorem
\ref{th:PCMS}  we have
$$
E^{X}(\De)\rh\otimes\ketbra{\xi}=E^{Y}(\De)\rh\otimes\ketbra{\xi}
$$
for any $\De\in\cB(\R)$.
Since $\Pi_{1}(\De)=V_{\xi}^{\da}E^{X}(\De)V_{\xi}$ and
$\Pi_{2}(\De)=V_{\xi}^{\da}E^{Y}(\De)V_{\xi}$ for all $\De$, where
$V_{\xi}\ps=\ps\otimes\xi$ for any $\ps\in\cH$.
Let $\ps\in\cH$.
We have
$
\Pi_{1}(\De)\rh\ps=V_{\xi}^{\da}E^{X}(\De)V_{\xi}\rh\ps
=V_{\xi}^{\da}E^{X}(\De)(\rh\ps\otimes\xi)
=V_{\xi}^{\da}E^{Y}(\De)(\rh\ps\otimes\xi)
=V_{\xi}^{\da}E^{X}(\De)V_{\xi}\rh\ps
=\Pi_{2}(\De)\rh\ps,
$
and by the standard argument we have
$
\Pi_{1}(f)\rh\ps=\Pi_{2}(f)\rh\ps
$
for any $f\in B(\R)$.
Since $\ps$ is arbitrary, we obtain \Eq{PCPOVM}.
\end{Proof}

\subsection{Perfect correlations between observables and POVMs}

For any observable $X$ and POVM $\Pi$,
we say that $X$ and $\Pi$ are {\em perfectly correlated} in
a state $\rh$, iff $E^{X}$ and $\Pi$ are perfectly correlated in $\rh$.
Now, we extend Theorem \ref{th:PCMS} to arbitrary pair of
an observable and a POVM.

\begin{Theorem}\label{th:PC-state3}
For any observable $X$, any POVM $\Pi$, and any
state $\rh\in\cS(\cH)$,
the following conditions are equivalent.

(i) $X$ and $\Pi$ are perfectly correlated in $\rh$.

(ii) $X$ and $\Pi$ are perfectly correlated in any state
$\si\in\cS(X,\rh)$.

(iii) $E^{X}(\De)\rh=\Pi(\De)\rh$ for any $\De\in\cB(\R)$.

(iv) $f(X)\rh=\Pi(f)\rh$ for any $f\in\B(\R)$.

(v) $f(X)P_{X,\rh}=\Pi(f)P_{X,\rh}$ for any $f\in\B(\R)$.
\end{Theorem}
\begin{Proof}
The implication (i)$\Then$(iv) follows from Theorem \ref{th:distance-POVM}.
The implication (iv)$\Then$(iii) is obvious.
The implication (iii)$\Then$(i) follows from the relations
\beqa
\Tr[E^{X}(\De)\Pi(\Ga)\rh]=
\Tr[E^{X}(\De)E^{X}(\Ga)\rh]=\Tr[E^{X}(\De\cap\Ga)\rh],
\eeqa
for any $\De,\Ga\in\cB(\R)$. Now, we shall show the implications
(i)$\Then$(v)$\Then$(ii)$\Then$(i).
Suppose that condition (i) holds.
Let $(\cK,\xi,L)$ be a Naimark-Holevo dilation of $\Pi$.
Then,  it is easy to see that 
%we have
%\beqa
%\bracket{E^{X}(\De)\ps',\Pi(\Ga)\ps}=
%\bracket{(E^{X}(\De)\ps')\otimes\xi,E^{L}(\Ga)(\ps\otimes\xi)}
%\eeqa
%for any $\ps,\ps'\in\cH$ and $\De,\Ga\in\cB(\R)$, so that
$(\cK,\xi,X\otimes I,L)$ is a joint dilation of $E^{X}$ and $\Pi$.
%Since $X$ and $\Pi$ are perfectly correlated in $\rh$, 
It follows from the assumption and 
Theorem \ref{th:distance-POVM} that $X\otimes I$
and $L$ are perfectly correlated in $\rh\otimes\ketbra{\xi}$, and hence
\beqa
f(X)\rh\ps\otimes \xi=f(L)(\rh\ps\otimes\xi)
\eeqa
for any $f\in B(\R)$ and $\ps\in\cH$.
Let $\ph\in\cH$.
Then, we have
$f(X)g(X)\rh\ps\otimes\xi=f(L)g(L)(\rh\ps\otimes\xi)
=f(L)(g(X)\rh\ps\otimes\xi)$
for any $f,g\in B(\R)$. 
%We have $P_{X,\rh}\ph\in\cC(X,\ps)$ and there is some $g\in B(\R)$ such
%that
%$P_{X,\ps}\ph=g(X)\ps$.
Thus, we have
\beqas
f(X)g(X)\rh\ps&=&V_{\xi}^{\da}(f(X)g(X)\rh\ps\otimes\xi)=
V_{\xi}^{\da}f(L)(g(X)\rh\ps\otimes\xi)\\
&=& V_{\xi}^{\da}f(L)V_{\xi}g(X)\rh\ps.
\eeqas
Since the vector of the form $g(X)\rh\ps$ with $g\in B(\R)$,
$\ps\in\cH$ spans $\cC(X,\rh)$, we obtain
$$
f(X)P_{X,\rh}= V_{\xi}^{\da}f(L)V_{\xi}P_{X,\rh}
$$
and hence, the implication (i)$\Then$(v) follows. 
Suppose that condition (v) holds.
Let $\si\in\cS(X,\rh)$ and $\De,\Ga\in\cB(\R)$.
Then, $P_{X,\rh}\si=\si$ and 
we have $E^{X}(\Ga)\si=\Pi(\Ga)\si$.
Thus,  
$$
\Tr[E^{X}(\De)\Pi(\Ga)\si]=
\Tr[E^{X}(\De)E^{X}(\Ga)\si]=
\Tr[E^{X}(\De\cap\Ga)\si].
$$
It follows that $X$ and $Y$ are perfectly correlated in $\si$,
and hence the implication (v)$\Then$(ii) follows.
Since the implication (ii)$\Then$(i) is obvious, the proof is completed.
\end{Proof}

\section{Perfect correlations in measurements}
\label{se:8}
\subsection{Quantum instruments and measuring processes}

A {\em measuring process} for $\cH$ is defined to be a quadruple
$(\cK,\xi,U,M)$ consisting of a separable
Hilbert space $\cK$, a state vector $\xi$ in $\cK$, 
a unitary operator $U$ on $\cH\otimes\cK$,
and an observable $M$ on $\cK$ \cite{84QC}.
It is a plausible hypothesis in the theory of measurement that
to any measuring apparatus $\bA(\bx)$ with output variable $\bx$
for a system $\bS$ described by a Hilbert space $\cH$, 
there corresponds to a measuring process $(\cK,\xi,U,M)$
such that $\cK$ describes the probe $\bP$ prepared in 
$\xi$ just before the measurement,
$U$ describes the time evolution of the composite system $\bS+\bP$
during the measuring interaction,
and that $M$ describes the meter observable to be actually observed
just after the measuring interaction 
\cite{84QC,BLM91,BK92,00MN,01OD,04URN}.
Then, the probability distribution of the output $\bx$
on input state $\rh$ is given by
\beqa
\Pr\{\bx\in\De\|\ \rh\}
=\Tr[(I\otimes E^{M}(\De))U(\rh\otimes\ketbra{\xi})U^{\da}],
\eeqa
and the conditional output state $\rh_{\{\bx\in\De\}}$ of
the apparatus on input state $\rh$ 
given the outcome $\bx\in\De$ is described by
\beqa
\rh_{\{\bx\in\De\}}
=
\frac{\Tr_{\cK}[(I\otimes E^{M}(\De))U(\rh\otimes\ketbra{\xi})U^{\da}]}
{\Tr[(I\otimes E^{M}(\De))U(\rh\otimes\ketbra{\xi})U^{\da}]},
\eeqa
where $\Tr_{\cK}$ stands for the partial trace over $\cK$.

Two measuring apparatuses $\bA(\bx),\bA(\by)$,
or corresponding measuring processes, 
 are called {\em
statistically equivalent} iff they have the same output distributions
and the same conditional output states on each input state, 
i.e., $\Pr\{\bx\in\De\|\ \rh\}=\Pr\{\by\in\De\|\ \rh\}$ and $\rh_{\{\bx\in\De\}}
=\rh_{\{\by\in\De\}}$ for all $\rh\in\cS(\cH)$\ and $\De\in\cB(\R)$.
The statistical equivalence classes of all the measuring processes
are characterized by completely positive map valued measures
as follows. 

Denote by $\ta c(\cH)$ the space of trace class operators on $\cH$
and by $\cL(\tc(\cH))$ the space of bounded linear transformations
on $\tc(\cH)$.  
A linear transformation $T\in\cL(\tc(\cH))$ is called {\em completely
positive} iff $T\otimes\id_{n}\in\cL(\tc(\cH\otimes\C^{n}))$ is a
positive transformation for any positive integer $n$. 
Denote by $\cC\cP(\tc(\cH))$ the space of completely positive maps
on $\tc(\cH)$.
An {\em instrument} is a countably additive normalized 
completely positive map valued measure from $\cB(\R)$ to $\cL(\tc(\cH))$,
i.e., a mapping $\cI:\cB(\R)\to\cC\cP(\tc(\cH))$ satisfying that
$\cI(\R)$ is trace-preserving and 
$\sum_{j=1}^{\infty}\cI(\De_{j})=\cI(\R)$ 
in the strong operator topology for any disjoint Borel sets 
$\De_{1},\De_{2},\ldots$
such that $\bigcup_{j}\De_{j}=\R$ \cite{84QC}.

For any instrument $\cI$ and state $\rh$,
the relation
\beqa
\mu_{\rh}^{\cI}(\De)=\Tr[\cI(\De)\rh]
\eeqa
defines a probability measure on $\cB(\R)$ called the {\em output
distribution} of $\cI$ on input state $\rh$,
and the state
\beqa
\frac{\cI(\De)\rh}{\Tr[\cI(\De)\rh]}
\eeqa
is called the {\em output state} of $\cI$ on input state $\rh$
given $\De$ \cite{DL70}.
The dual map of $\cI(\De)$ is the linear transformation
$\cI(\De)^{*}$ on $\cL(\cH)$ defined by
\beqa
\Tr[(\cI(\De)^{*}A)\rh]=\Tr[A\cI(\De)\rh]
\eeqa
for all $A\in\cL(\cH)$, $\rh\in\tc(\cH)$, and $\De\in\cB(\R)$.
Then, $\cI(\De)^{*}$ is a normal completely positive map
on $\cL(\cH)$ \cite{Tak79} and especially $\cI(\R)^{*}$ is unit-preserving.
The relation 
\beqa
\Pi^{\cI}(\De)=\cI(\De)^{*}I
\eeqa
where $\De\in\cB(\R)$ defines a POVM,
called the {\em POVM} of $\cI$, which satisfies 
\beqa
\mu_{\rh}^{\cI}(\De)=\Tr[\Pi^{\cI}(\De)\rh]
\eeqa
for all $\De\in\cB(\R)$ and $\rh\in\cS(\cH)$.

For any measuring process $\M=(\cK,\xi,U,M)$, the relation 
\beqa
\cI_{\M}(\De)\rh
=\Tr_{\cK}[(I\otimes E^{M}(\De))U(\rh\otimes\ketbra{\xi})U^{\da}],
\eeqa 
where $\rh\in\cS(\cH)$ and $\De\in\cB(\R)$, defines an instrument
$\cI_{\M}$, called the {\em instrument} of $\M$.
Then, the POVM of $\cI$ is called the {\em POVM} of $\M$ and
denoted by $\Pi_{\M}$.  We have
\beqa
\Pi_{\M}(\De)=\cI_{\M}(\De)^{*}I
=
\Tr_{\cK}[U^{\da}(I\otimes E^{M}(\De))U(I\otimes\ketbra{\xi})]
\eeqa
for all $\De\in\cB(\R)$.
For all $\rh\in\cS(\cH)$ and $\De\in\cB(\R)$, we have
\beqa
\Pr\{\bx\in\De\|\ \rh\}
&=&\Tr[\cI_{\M}(\De)\rh]=\Tr[\Pi_{\M}(\De)\rh]
\eeqa
and 
\beqa
\rh_{\{\bx\in\De\}}&=&\frac{\cI(\De)\rh}{\Tr[\cI(\De)\rh]},
\eeqa
provided $\Tr[\cI(\De)\rh]>0$.
Thus, two measuring processes are statistically equivalent
if and only if they have the same instrument.

Conversely, it has been proved in Ref.~\cite{84QC} that
for any instrument $\cI$, there exists a measuring process
$\M=(\cK,\xi,U,M)$ such that $\cI=\cI_{\M}$.
Thus, every instrument corresponds at least one measuring
process, and therefore the instruments are
in one-to-one correspondence with the statistical
equivalence classes of measuring processes.

\subsection{Precise measurements of observables}

Once the notion of measurement has been fully generalized by
the notion of instruments, a fundamental problem is to recover the
conventional notion of measurements of observables in this 
general formulation.  In what follows, we shall give an answer
to this problem in the light of the notion of quantum
perfect correlations.

According to a fundamental postulate of quantum mechanics, 
if an apparatus $\bA(\bx)$ measures an observable $A$ in a state $\rh$, 
the probability distribution of the output $\bx$ on input state $\rh$ should
satisfy the {\em Born statistical formula} (BSF) 
\beqa\label{eq:BSF}
\Pr\{\bx\in\De\|\rh\}=\Tr[E^{A}(\De)\rh],
\eeqa
where $\De\in\cB(\R)$.
From the above it is tempting to say that an apparatus $\bA(\bx)$ 
measures observable $A$ in state $\rh$, iff it satisfies the BSF \eq{BSF}.
However, to reproduce the probability
distribution of observable $A$ in state $\rh$ is a necessary but not
sufficient condition for the apparatus $\bA(\bx)$ to measure $A$ in $\rh$.
For example, suppose $\cH=\cK\otimes\cK$, $\rh=\si\otimes\si$,
and $A=X\otimes I$ and $B=I\otimes X$ for some Hilbert space $\cK$,
a state $\si$ of $\cK$, and an observable $X$ of $\cK$.
In this case, we have
$\Tr[E^{A}(\De)\rh]=\Tr[E^{B}(\De)\rh]=\Tr[E^{X}(\De)\si]$,
so that every apparatus $\bA(\by)$ measuring $B$ in state $\rh$ 
also satisfies the BSF  for $A$ in $\rh$.
However, we cannot consider that the apparatus $\bA(\by)$ 
measures $A$ in $\rh$ as well as $B$ in $\rh$.
Since $A$ and $B$ are independent observables in the separated 
subsystems,
so that another apparatus $\bA(\bx)$ may simultaneously measure $A$ 
and may obtain a different outcome of the $A$ measurement.  
In this case, we can say that the apparatus $\bA(\bx)$ measures $A$ 
but the apparatus $\bA(\by)$ does not.

In order to find a satisfactory condition to ensure that a given 
instrument $\cI$ measures $A$ in $\rh$, let us consider a measuring 
process $\M=(\cK,\xi,U,M)$ of $\cI$.
Suppose that we measure $A$ at time $t$ at which the system $\bS$ 
described by Hilbert space $\cH$ is 
in state $\rh$ and that the measuring interaction turns on from time
$t$ to $t+\De t$.  In the Heisenberg picture with the original
state $\rh\otimes\ketbra{\xi}$, we write $A(t)=A\otimes I$,
$A(t+\De t)=U^{\da}(A\otimes I)U$, $M(t)=I\otimes M$,
and $M(t+\De t)=U^{\da}(I\otimes M)U$.
Then, in order to measure $A(t)$, this measurement actually
measures $M(t+\De t)$, so that observables $A(t)$ and $M(t+\De t)$ 
should be perfectly correlated in the original state $\rh\otimes\ketbra{\xi}$.

In the previous example, it is concluded that the meter observable of
$\bA(\by)$ after the measuring interaction, $M(t+\De t)$, 
cannot be perfectly correlated with the observable $A$ 
before the interaction, $A(t)$.
In fact, $M(t+\De t)$ should be 
perfectly correlated with the observable $B$ before the interaction, $B(t)$,
while $A(t)$ and $B(t)$ are not perfectly correlated before the interaction.
It follows from the transitivity of perfect correlations that 
$A(t)$ and $M(t+\De t)$ cannot be perfectly correlated.

It is also clear that given two ``meter'' observables $M_{1}$
and $M_{2}$ in the external system described by a Hilbert
space $\cK$ and given the original state $\rh\otimes\ketbra{\xi}$
of $\cH\otimes\cK$ at time $t$, if both the pair of 
$A(t)$ and $M_{1}(t+\De t_{1})$ and the pair of 
$A(t)$ and $M_{2}(t+\De t_{2})$ are perfectly correlated in the
original state, 
then we can conclude that both meters give the concordant outcome
from the transitivity of perfect correlations.

According to the above consideration, it is natural to say that 
a measuring process $\M=(\cK,\xi,U,M)$ {\em precisely measures} 
an observable
$A$ on input state $\rh$ iff the observable $A\otimes I$ and 
$U^{\da}(I\otimes M)U$ are perfectly correlated in the state 
$\rh\otimes\ketbra{\xi}$,
and that an instrument $\cI$ {\em precisely measures} an observable
$A$ on input state $\rh$ iff every measuring process $\M$ for $\cI$
precisely measures $A$ on input state $\rh$.
In the above, the adverb ``precisely'' is used to distinguish this case 
from any approximate measurements of the same observable.
 
The following theorem shows that whether the measuring process $\M$
precisely measures $A$ on $\rh$ is determined solely 
by the corresponding POVM.

\begin{Theorem}\label{th:precise_measurements}
A measuring process $\M=(\cK,\xi,U,M)$ 
precisely measures an observable $A$ in a state $\rh$ if and only if
the POVM of $\M$ is perfectly correlated with the observable
$A$ in the state $\rh$.
\end{Theorem}
\begin{Proof}
The assertion follows immediately from the relations
\beqas
\Tr[(E^{A}(\De)\otimes I)U^{\da}(I\otimes E^{M}(\Ga))U(\rh\otimes\ketbra{\xi})]
&=&
\Tr[E^{A}(\De)\Pi_{\bx}(\Ga)\rh],\\
\Tr[(E^{A}(\De)\otimes I)(\rh\otimes\ketbra{\xi})]
&=&
\Tr[E^{A}(\De)\rh].
\eeqas
\end{Proof}

The following theorem characterizes, up to statistical equivalence,
the precise measurements of an observable in a given state.

\begin{Theorem}
For any instrument $\cI$ with POVM $\Pi^{\cI}$,
any observable $A$, and any state $\rh$,
the following conditions are all equivalent.

(i) $\cI$ precisely measures $A$ in $\rh$.

(ii) $\Pi^{\cI}$ is perfectly correlated to $A$ in $\rh$.

(iii) $\Pi^{\cI}$ is perfectly correlated to $A$ in any
state $\si\in\cS(A,\rh)$.

(iv) $\cI$ satisfies the BSF for $A$ in any state $\si\in\cS(A,\rh)$.

(v) $\Pi^{\cI}(\De)\si=E^{A}(\De)\si$ for any $\si\in\cS(A,\rh)$ and 
$\De\in\cB(\R)$.

(vi) $\Pi^{\cI}(\De)P_{A,\rh}=E^{A}(\De)P_{A,\rh}$ for any
$\De\in\cB(\R)$.
\end{Theorem}
\begin{Proof}
The assertion follows easily from Theorems \ref{th:PC-state3} and
\ref{th:precise_measurements}.
\end{Proof}

In the conventional interpretation of instruments proposed
by Davies and Lewis \cite{DL70}, an instrument $\cI$ is 
considered to precisely measure $A$ in {\em every} state $\rh$
iff it satisfies the BSF for $A$ in every state $\rh$.
Since the BSF for $A$ in a given state $\rh$ does not ensure
that the instrument $\cI$ precisely measures $A$ in $\rh$,
the above hypothesis lacks an immediate justification
in the sense that it is not immediately clear whether this hypothesis
excludes the ambiguity of the simultaneous meter readings 
of the same observable.
However, this hypothesis has been finally justified 
by the above theorem, which concludes that
$\cI$ precisely measures $A$ in every state $\rh$ 
if and only if $\cI$ satisfies the BSF for $A$ in every state.

\subsection{von Neumann's  model of repeatable measurement}

It was shown by von Neumann \cite{vN32} 
that a repeatable measurement of an
observable 
\beqa\label{eq:A}
A=\sum_{n}a_{n}\ketbra{\ph_{n}}
\eeqa
 on $\cH$
with eigenvalues $a_{1},a_{2},\ldots$ and orthonormal basis
of eigenvectors
$\ph_{1},\ph_{2},\ldots$ can be realized by a unitary operator 
$U$ on the tensor product $\cH\otimes\cK$  with another separable
Hilbert space $\cK$ with orthonormal basis $\{\xi_{n}\}$ 
such that
\beqa\label{eq:U}
U(\ph_{n}\otimes\xi)=\al_{n}\ph_{n}\otimes\xi_{n},
\eeqa
where $\xi$ is an arbitrary state vector in $\cK$,
and $\al_{n}$ is an arbitrary phase factor, i.e., $|\al_{n}|=1$, for all $n$.
Let 
\beqa\label{eq:B}
M=\sum_{n}a_{n}\ketbra{\xi_{n}}
\eeqa
be an observable on $\cK$ called the meter.
von Neumann's model defines an apparatus $\bA(\bx)$ with
measuring process $(\cK,\xi,U,M)$.

Let us suppose that the initial state of the system is given by an arbitrary
state vector $\ps=\sum_{n}\sqrt{p_{n}}\ph_{n}$.
Then, it follows from the linearity of $U$ we have
\beqa\label{eq:U-sp}
U(\ps\otimes\xi)=\sum_{n}\sqrt{p_{n}}\ph_{n}\otimes\xi_{n}.
\eeqa
The conventional explanation as to why this transformation can be
regarded as a measurement is as follows; symbols are adapted to
the present context in the quote below.  ``In the state \eq{U-sp},
obtained by the measurement, there is a statistical correlation between
the state of the object and that of the apparatus: the simultaneous 
measurement on the system---object-plus-apparatus---of the two quantities,
one of which is the originally measured quantity of the object and the
second the position of the pointer of the apparatus, always leads to concordant
results.  As a result, one of these measurements is unnecessary:  The state
of the object can be ascertained by an observation on the apparatus.
This is a consequence of the special form of the state vector \eq{U-sp},
on not containing any $\ph_{m}\otimes\xi_{n}$ term with $n\not=m$
\cite{Wig63}.''
``The equations of motion permit the description of the process whereby
the state of the object is mirrored by the state of an apparatus.
The problem of a measurement on the object is thereby transformed into
the problem of an observation on the apparatus \cite{Wig63}.''

The above explanation correctly points out the existence of the 
statistical correlation between the measured observable $A$ and
the meter observable $M$ in the state \eq{U-sp}.  However,
this is not the statistical correlation between the measured observable
before the interaction and the meter observable after the interaction,
but that between those observables after the interaction.
Thus, the above statistical correlation does not even ensure that
the probability distribution of the measured observable before the
interaction is reproduced by the observation of the meter observable
after the interaction.

The role of the measuring interaction described by
$U$ should be to make the following two correlations:
(i) the correlation between the measured observable
$A$ {\em before} the interaction and the meter 
$M$ {\em after} the interaction, and (ii) the correlation between
the meter $M$ {\em after} the interaction and 
the measured observable $A$ {\em after} the interaction.
The first correlation is required by the {\em value reproducing requirement}
that the interaction transfers the value of the measured observable
$A$ before the interaction to the value of the meter $M$ after
the interaction.
The second correlation is required by the {\em repeatability hypothesis}
that if the meter observable $M$ has the value $a_{n}$ after the
interaction, then the observable $A$ also have the same value $a_{n}$
after the interaction so that the second measurement of $A$ after the
interaction reproduce the same value of the meter of the first
measurement of $A$.  

Now, we shall show that those requirements are actually satisfied.
Let $\et_{0},\et_{1},\ldots$ be an orthonormal basis of $\cH$ such that 
$\et_{0}=\xi$, namely an orthonormal basis extending $\{\xi\}$.
Let $\Ps_{n,m}$ be a unit vector in $\cH$ defined by
$\Ps_{n,m}=U^{\da}(\ph_{n}\otimes\xi_{m})$ for any $n,m$.
Then, we have $\Ps_{n,n}=\ph_{n}\otimes\xi$ and the 
family $\{\Ps_{n,m}\}$ is an orthonormal basis of $\cH$.
By simple calculations, we have
\beqa
A\otimes I
&=&A\otimes\ketbra{\xi}
+\sum_{m\not=0}A\otimes\ketbra{\et_{m}},\\ 
U^{\da}(A\otimes I)U
&=&A\otimes\ketbra{\xi}
+\sum_{n\not=m}a_{n}\ketbra{\Ps_{n,m}},\\
U^{\da}(I\otimes M)U
&=&A\otimes\ketbra{\xi}
+\sum_{n\not=m}a_{m}\ketbra{\Ps_{n,m}},
\eeqa
where $\sum_{n\not=m}$ stands for the summation over all
$n,m$ with $n\not=m$.  By the above relations it is now obvious that
$A\otimes I=U^{\da}(A\otimes I)U=U^{\da}(I\otimes M)U$
on their common invariant subspace $\cH\otimes [\xi]$,
so that those three observables are perfectly correlated in
the state $\ps\otimes\xi$ for every state vector $\ps$ in $\cH$.
Therefore, von Neumann's model $(\cK,\xi,U,M)$ satisfies both
the the value reproducing requirement and the repeatability
hypothesis.

The following theorem characterizes the unitary operators
that fulfil the above two requirements.

\begin{Theorem}
Let $\{\ph_{n}\}$ and $\{\xi_{n}\}$ be orthonormal bases of $\cH$
and $\cK$, respectively, and the observables $A$ and $B$ be given by
\Eq{A} and \Eq{B}, respectively.
Then,  a unitary operator $U$ on $\cH\otimes\cK$ and a state vector 
$\xi\in \cK$ satisfy \Eq{U} if and only if (i)
$A\otimes I$ and $U^{\da}(I\otimes B)U$ are perfectly correlated in
$\ps\otimes\xi$ and that (ii)
$U^{\da}(A\otimes I)U$ and $U^{\da}(I\otimes B)U$ are perfectly correlated
in $\ps\otimes\xi$ for every state vector $\ps\in\cH$.
\end{Theorem}
\begin{Proof}
Suppose that $U$ and $\xi$ satisfy \Eq{U}.
Without any loss of generality we assume
$U(\ph_{n}\otimes\xi)=\ph_{n}\otimes\xi_{n}$ for all $n$;
otherwise, we can replace $\xi_{n}$ by $\al_{n}\xi_{n}$ 
without changing $B$.
Let $\ps=\sum_{n}c_{n}\ph_{n}$.
By linearity of $U$ we have
$U(\ps\otimes\xi)=\sum_{n}c_{n}\ph_{n}\otimes\xi_{n}$.
Thus, it follows 
from the argument on the entangled state given \Eq{entangled},
$A\otimes I$ and $I\otimes B$ are perfectly correlated in $U(\ps\otimes\xi)$.
By Theorem \ref{th:equivalence-3}, 
$U^{\da}(A\otimes I)U$ and $U^{\da}(I\otimes B)U$ 
are perfectly correlated in $\ps\otimes\xi$.
Thus, condition (ii) holds.
Let $\{\et_{n}\}$ be an orthonormal basis of $\cK$ 
such that $\et_{1}=\xi$.
Then, we have
$$
U(E^{A}(a_{n})\otimes I)(\ps\otimes\xi)
=
U(\ketbra{\ph_{n}}\otimes I)\sum_{j}c_{j}\ph_{j}\otimes\xi
=c_{n}U(\ph_{n}\otimes\xi)
=c_{n}\ph_{n}\otimes\xi_{n},
$$
and 
$$
(I\otimes E^{B}(a_{m}))U(\ps\otimes\xi)
=
(I\otimes\ketbra{\xi_{m}})\sum_{j}c_{j}\ph_{j}\otimes\xi_{j}
=
c_{m}\ph_{m}\otimes\xi_{m}.
$$
Thus, we have
\beqas
\bracket{(E^{A}(a_{n})\otimes I)(\ps\otimes\xi),
U^{\da}(I\otimes E^{B}(a_{m}))U(\ps\otimes\xi)}
=
c_{n}^{*}c_{m}\de_{n,m},
\eeqas
and this shows that $A\otimes I$ and $U^{\da}(I\otimes B)U$
are perfectly correlated in $\ps\otimes\xi$.
Thus, we have proved the necessity of conditions (i) and (ii).
Conversely, suppose that conditions (i) and (ii) hold.
Let $\ps=\ph_{n}$.
Since $A\otimes I$ and $U^{\da}(I\otimes B)U$ are perfectly
correlated in $\ps\otimes\xi$, we have $$
\bracket{(I\otimes E^{B}(a_{n}))U(\ph_{n}\otimes \xi),U(\ph_{n}\otimes\xi)}
=
\bracket{(E^{A}(a_{n})\otimes I)(\ph_{n}\otimes\xi),(\ph_{n}\otimes \xi)} =1.$$
Thus, $U(\ph_{n}\otimes\xi)=\et_{n}\otimes\xi_{n}$ for some state vector
$\et_{n}$.
Since $A\otimes I$ and $I\otimes B$ are perfectly
correlated in $U(\ps\otimes\xi)$, we have 
$$
\bracket{E^{A}(a_{n})\et_{n},\et_{n}}=
\bracket{(E^{A}(a_{n})\otimes I)(\et_{n}\otimes\xi_{n}),(\et_{n}\otimes\xi_{n})}
=
\bracket{(I\otimes E^{B}(a_{n}))(\et_{n}\otimes\xi_{n}),
(\et_{n}\otimes\xi_{n})}=1.
$$
Thus, $\ketbra{\et_{n}}=\ketbra{\ph_{n}}$, so that $U$ and $\xi$ satisfy
\Eq{U}.
\end{Proof}

Now, we return  to von Neumann's measurement model described by
\Eq{U}.
The measurement is said to satisfy the {\em nondemolition condition},
iff the measured observable
is not disturbed by the measuring interaction, so that $A\otimes I$ and
$U^{\da}(A\otimes I)U$ is perfectly correlated in $\ps\otimes\xi$.
As we have shown in Theorem \ref{th:transitivity} perfect correlations are
transitive.  Thus, the perfect correlation between 
$A\otimes I$ and $U^{\da}(I\otimes B)U$ and that between 
$U^{\da}(A\otimes I)U$ and $U^{\da}(I\otimes B)U$ implies
the perfect correlation between $A\otimes I$ and
$U^{\da}(A\otimes I)U$. 
In the same way, we will be able to explain that two out of three conditions, 
(i) the valued reproducing condition, (ii) the repeatability hypothesis, and (iii) the
nondemolition condition, imply the other one, as straightforward consequence 
of the transitivity of perfect correlations.

\section{Concluding remarks}\label{se:9}

Let $X,Y$ be a pair of (discrete) observables and $\ps$ a state.  
Consider the following conditions.

(i) (Equi-valuedness) 
No joint measurements of $X$ and $Y$ in $\ps$, if
any, give different  values, i.e., 
$$
\bracket{E^{X}(\De)\ps,E^{Y}(\Ga)\ps}=0
$$ 
if $\De\cap\Ga=\emptyset$.

(ii) (Reproducibility)
Successive projective measurements of $X$ and $Y$ in $\ps$
always give the same value irrespective of the order of measurements, i.e, 
$$
\sum_{y\in\Ga}\|E^{X}(\De)E^{Y}(\{y\})\ps\|^{2}=
\sum_{x\in\De}\|E^{Y}(\Ga)E^{X}(\{x\})\ps\|^{2}=0
$$
if $\De\cap\Ga=\emptyset$.

(iii)  (Zero difference) The difference $X-Y$ has the definite value zero in
$\ps$. i.e.,
$$
(X-Y)\ps=0.
$$

(iv)  (Identical distributivity)
Independent measurements of $X$ and $Y$ in $\ps$ have the identical
output probability distribution, i.e.,
$$
\|E^{X}(\De)\ps\|^{2}=\|E^{Y}(\De)\ps\|^{2}
$$
for any $\De\in\cB(\R)$.

In this paper, we have shown the following logical relations
among the above conditions.
The following implications holds: 
(i)$\Iff$(ii), (i)$\Then$(iii), (i)$\Then$(iv).  
However, none of the implications (iii)$\Then$(i), (iii)$\Then$(iv),
(iv)$\Then$(i), and (iv)$\Then$(iii) hold.
If $X$ and $Y$ commute, (i)$\Iff$(iii) and (iii)$\Then$(iv) holds, 
but (iv)$\Then$(iii) still does not hold.
In order to clarify the mutual relations, we have considered the notion of
the cyclic subspace $\cC(X,\ps)$ or $\cC(Y,\ps)$ and required
conditions (iii) and (iv) to be satisfied by any state $\ph$ in $\cC(X,\ps)$ 
or $\cC(Y,\ps)$, as follows.

(iii)' $(X-Y)\ph=0$ for any $\ph\in\cC(X,\ps)$.

(iv)' $\|E^{X}(\De)\ph\|^{2}=\|E^{Y}(\De)\ph\|^{2}
$ for any $\De\in\cB(\R)$ and any $\ph\in\cC(X,\ps)$.

Then, we have shown that all the conditions (i), (ii), (iii)', and (iv)'
are mutually equivalent.  
According to this, we have proposed and justified 
to say that $X$ and $Y$ are perfectly correlated in $\ps$
iff one of the above equivalent conditions is satisfied.

We have also given an appropriate generalizations of the above considerations
to arbitrary observables $X,Y$ and arbitrary state $\rh$.

We have shown that so defined relation $X\equiv_{\rh}Y$
meaning $X$ and $Y$ are perfectly correlated in $\rh$ is
an equivalence relation on all the observables.  In particular,
if $X$ and $Y$ are perfectly correlated as well as $Y$ and $Z$,
we can conclude that so are $X$ and $Z$.
This suggests that perfectly correlated observables can be
interpreted to have the same value that can be realized
by joint measurements of them, even though the quantum state
determines it only randomly.

The above interpretation has given a new insight on the state
dependent definition of precise measurements of observables.
Even though
the outcome of a measurement might be used to infer what is
the state before or after the measurement as in quantum state
estimation or quantum state reduction, this inference cannot
be done without appealing to the fact that any measurement measures
some observable in the sense of the Born rule; 
recall that even a POVM measurement corresponds 
to a measurement of an observable in a larger system and as such a
mathematical POVM can be identified with a real experiment.  
Thus, the most fundamental question in measurement theory
is the one as to what observable is
(precisely) measured by a given apparatus.  

Conventionally,
this question has been answered only in a state independent
manner as follows: The apparatus measures an observable $X$ 
if and only if the probability reproducing condition (PRC) is satisfied
for {\em any} input state, where the PRC requires that
the output probability
distribution reproduce the theoretical probability distribution predicted by
the Born rule.
However, the justification of the above definition has not been clear,
since the probability reproducing condition 
for {\em a given} input state does not imply that
the measurement is precise in that state.
In this respect, our result has successfully justified the conventional
definition in that we have given a definition of a precise
measurement in a given state and showed that the conventional
definition indeed requires the measurement is precise in any input
state. 

The state dependent definition is not only a pedantic justification of
the conventional approach.
In fact, some measuring
apparatus in a laboratory can accept only a small class of states
from the whole Hilbert space of the state vectors. 
For instance, every microscope cannot measure the position
of a particle outside of the scope.  Thus, the
experimenter should have a criterion to judge whether or not the
apparatus measures the given observable depending on the
input state.  Such a criterion was not even discussed in measurement
theory before the present investigation.

\section*{Acknowledgments}
This work was supported by the 
Strategic Information and Communications R\&D Promotion Scheme
of the MPHPT of Japan,  by the CREST
project of the JST, and by the Grant-in-Aid for Scientific Research of
the JSPS.

\end{document}